\documentclass[10pt,flushrt,preprint,longnamesfirst]{aastex}
\usepackage{epsfig}
\usepackage{xspace}

\slugcomment{Accepted July 8, 2003 for publication in the {\em AJ}}
\shorttitle{Light Echo Observability}
\shortauthors{Sugerman}

\begin{document}
\title{Observability of Scattered-Light Echoes Around Variable Stars
and Cataclysmic Events}

\author{Ben E.\ K.\ Sugerman}
\affil{Department of Astronomy, Columbia University,
  New York, NY 10027}
\email{ben@astro.columbia.edu}

\begin{abstract}
Scattered light echoes from variable and cataclysmic stars offer one
of the most effective means to probe the structure and composition of
circumtellar and interstellar media.  I build a simple model of
light-echo surface brightness by considering the source spectrum, and
the dust density, geometry, and scattering efficiency.  I use this
model to investigate whether echoes should be observable around short
and long-period giants, cataclysmic variables, and supernovae.  Only
supernovae are expected to illuminate material on both circumstellar
and interstellar scales.  Giant and post-AGB stars (e.g.\ Cepheids and
Miras) with high mass-loss rates can produce observable echoes within
their circumstellar envelopes.  Echoes from novae and dwarf novae are
probably detectable only from nearby material, and only in
unusually dense gas.  I present characteristic exposure times to image
such echoes over a wide range of wavelengths for ground-based and {\em
Hubble Space Telescope} observations.  I apply these results to
analyze the dust properties of the recently-reported echoes around
SN~1993J, finding the dust in M81 to have a grain-size distribution
and chemical composition consistent with Galactic dust.  Optimal
observing strategies for echo detection are also discussed. 
\end{abstract}

\keywords{ 
circumstellar matter ---
ISM: dust ---
stars: supernova (SN~1993J) ---
stars: variables -- 
techniques: photometric 
}

\newpage

\section{INTRODUCTION \label{sec-intro}}

Since \citet{Cou39} proposed the theory  of ``aur\'eoles lumineuses''
to explain the otherwise superluminal sources surrounding Nova Persei
1901 \citep{Rit01}, scattered
light echoes have offered one of the most effective means to probe
circumstellar and interstellar structure.   When the light pulse from
some varying object is scattered by dust into the line of sight, an
observable echo is produced, provided the pulse is sufficiently
luminous and the dust sufficiently dense.  An echo observed a given time
after the pulse must lie on the locus of points equidistant in
light travel from the source and observer, i.e.\ an ellipsoid with
known foci.  This simple geometry directly yields the three-dimensional
position of an echo, uncertain only by the assumed distance to the source.

A great challenge to any observer is determining the distances
of objects of interest along the line of sight.  Whereas this normally
requires a variety of positional and absorption-velocity information
as well as a 
geometric model, the relative line-of-sight positions of light echoes
are unambiguous, 
and can be determined with as little as a single image.  Given a
reliable distance determination to the source, the absolute position
of an echo and the precise three-dimensional structure of the
surrounding dust are known.  Conversely, light echoes can be employed
to determine the distance to structures with independently-known
geometry.  Finally, as the spectrum of an
echo is determined by the scattering properties of the dust, light
echoes constrain the grain-size distribution, density, and composition
of dust in the circumstellar (CSM) and interstellar medium (ISM),
provided the input spectrum is known.  

Light echoes have been considered in connection with gamma-ray bursts
\citep{Rei01,BH92} and active galactic nuclei \citep{deD98}, as distance
indicators to galaxy clusters
\citep{Mil87,Kat87,BM89}, and for probing the baryon density of the
universe with high-$z$ sources \citep{Sho97}.  If nearly-90\degr\ 
scattering occurs, the positions of maximally-polarized echoes yield
geometric distance measurements to the echo source \citep{Spa94,Spa96}.
The influence of echoes on the
light curves of type II supernovae (SNe) has been discussed by e.g.,\
\citet{DiC02}, \citet{RS00}, \citet{Fil95}, \citet{Chu92}, and
\citet{Mac87}.  The detection of echoes has been investigated around
SNe \citep{Wri80,Che86,Sch87a,Sch87b,EC89,Spa94,Mas00}, novae \citep{Geh88},
and flaring stars \citep{Arg74,Bro92,Gai94}.

In spite of the wealth of information contained in echoes, light echo
astronomy is still a nascent field.  This is due in part to the
relative difficulty of directly detecting them.  Foremost, the
brightest echo-producing objects (SNe) are also the rarest, and only a
limited number of recent candidates are close enough to produce
resolved echoes.  Targeted searches around supernovae have been
conducted by \citet{vdB65a,vdB65b,vdB66} and \citet{Bof99}.  The
latter search found 16 echo candidates out of a study of 65 historic
SNe in 38 galaxies.  Although no conclusive detections were made,
echoes could be discovered around these candidates in follow-up
observations.  Novae are the next likely class of targets, and of the
searches conducted by \citet{vdB77} and \citet{Sch88}, no echoes were
detected.  To date, echoes have actually been resolved around only a
handful of objects.  These include Nova (N) Per 1901
\citep{Cou39,Swe48}, N 1975 Cyg \citep{BE85}, N V838 Mon
\citep{Mun02,Bon03}, SN~1987A (see below), SN~1991T \citep{Spa94,Spa99},
SN~1993J \citep{Liu02,SC02}, and SN~1998bu \citep{Cap01}.  
Variability in the circumstellar material around the Cepheid variable
RS Pup \citep{Hav72,MEB85} and the biplolar nebula OH 231.8+4.2 \citep[the
Rotten Egg nebula;][]{Kas92} may also be interpreted as light echoes from
the central variable star.

SN~1987A in the Large Magellanic Cloud (LMC) is the nearest known SN
in the last 400 years.  Due to its proximity, echoes around SN~1987A
are the most widely discussed and extensively studied.  It is beyond
the scope of this article to give a full accounting of the
references--there are at least 15 publications discussing the
detection of echoes, not including regularly cited circulars and
conference proceedings.  These echoes have been used to map the
surrounding circumstellar (\citealt{Cro91}; \citealt{Cro95}; Sugerman
et al., in preparation) and interstellar 
\citep{Xu94,Xu95,Xu99} material, thereby probing the progenitor's
mass-loss history, its location within the LMC, as well as the
structure and history of the associated stars and gas.

The low detection rate from intentional searches and small
number of known echoes suggests that
they are especially difficult to image, and may have
discouraged interest in recent years.   One may also speculate
whether limitations in resolution and detector sensitivity, as well as
unoptimized observing strategies, have contributed to these disappointing
results and to the limited application of this technique.  Under what
circumstances can one hope to succesfully image light echoes?  Are
there better observational techniques or more likely candidates to
study?  How can one maximize the information content of those observed?  

In this article, I discuss the observability of scattered-light echoes
around a variety of variable sources, and present an optimized
observing method, which has already been proven successful.
Formalisms for dust scattering are developed in \S\ref{sec-formal}, in
which I build the components for predicting light-echo surface
brightness in \S\ref{sec-model}.  Dust scattering properties are
presented in \S\ref{model-dust}, candidate variable stars for
producing echoes in \S\ref{model-spec}, and input constraints in
\S\ref{model-const}.  These components are used in \S\ref{model-pred}
to compute the exposure times to image light echoes using ground-based
telescopes, and the Advanced Camera for Surveys (ACS) and Space
Telescope Imaging Spectrograph (STIS) aboard {\em Hubble Space
Telescope} ({\em HST}).  I apply the model to echoes recently
discovered around SN~1993J in \S\ref{sec-93J}, and discuss an optimal
observing strategy in the conclusions (\S\ref{sec-concl}).

\section{FORMALISM \label{sec-formal}}

\subsection{Light Echo Surface Brightness \label{form-sb}}

\begin{figure}\centering
\epsfig{file=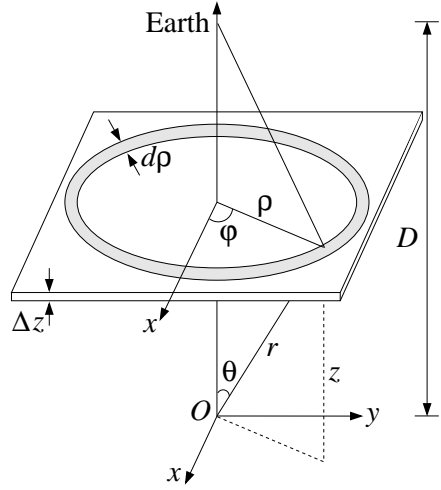,angle=0,width=0.45\linewidth}
\caption{Echo geometry used in this paper.  Note that the distances
  are not to scale, and $D$ is much larger than $z$. Adapted from
  \citet{Xu94}. 
 \label{echogeom}}
\end{figure}

Consider the geometry shown in Figure \ref{echogeom} \citep[adapted
from][]{Xu94}, in which an object $O$ at distance $D$ from the
observer emits a pulse of light of duration $\tau$ at time
$\tilde{t}$.  This light echoes off dust located at position
{\boldmath $r$} from $O$, and reaches the observer at time $t$.  This
formalism was first developed to discuss light echoes from supernovae
\citep{EC88,EC89}, for which $\tilde{t}$ is measured from the time of
explosion, and $t$ is measured from the arrival of first visual light.
For the more general case of any variable source, one can consider
$\tilde{t}$ to be the time at which maximum light was emitted from the
source, and $t$ is the time an echo is observed as measured from the
arrival of maximum light.

Let us treat the dust as a thin planar sheet of thickness $\Delta z$;
this approximation is only for computational convenience but will not
effect the generality of the result.  The light scattered at
{\boldmath $r$} lies on an ellipsoid with $O$ and the observer at the
foci.  Since $D$ is generally much greater than $r$, the echo depth
$z$ along the line-of-sight can be approximated by the parabola
\citep{Cou39}
\begin{equation} \label{sb1}
 z=\frac{\rho^2}{2ct}-\frac{ct}{2}
\end{equation}
where $\rho=r\sin{\theta}$ is the distance of the echo from the
line-of-sight in the plane of the sky, and $\theta$ is the scattering
angle.  Since the light pulse and dust
have finite widths, the echo will have a width $d\rho$ (\S\ref{form-width}).

The flux scattered off one dust grain of radius $a$ at position
{\boldmath $r$} is \citep{Che86,EC89}
\begin{equation} \label{sb2}
 dF_{\rm SC}(\lambda,t,\mbox{\boldmath $r$},a)= 
 \frac{Q_{\rm SC}(\lambda,a) \sigma_g L(\lambda,\tilde{t})\Phi(\mu,\lambda,a)}
 {16\pi^2 D^2r^2}
\end{equation}
where 
$Q_{\rm SC}$ is the grain scattering efficiency;
$\sigma_g=\pi a^2$ is the grain cross section;
$L(\lambda,\tilde{t})$ is the luminosity at $\lambda$ and $\tilde{t}$;
and $\Phi$ is the scattering phase function for a given
$\mu=\cos{\theta}=1-c(t-\tilde{t})/r$.  I adopt the \citet{HG41} phase
function 
\begin{equation} \label{sb3}
 \Phi(\mu,\lambda,a)=\frac{1-g^2(\lambda,a)}
 {\left[1+g^2(\lambda,a)-2g(\lambda,a)\mu \right]^{3/2}}
\end{equation}
with $g(\lambda,a)$ measuring the degree of forward scattering for a
given grain.  The total flux $F_{\rm SC}$ from a single scattering is
found by multiplying equation (\ref{sb2}) by the dust density 
$n_d(\mbox{\boldmath $r$},a)$ and integrating over the scattering
volume and all grain sizes.  I restrict this work to single-scattering
events, and refer the reader to \citet{Che86} for a discussion of
multiple-scatterings, which are only of significance when the optical
depth of the scattering material $\gtrsim 0.3$.
  
Proceeding as in \citet{Xu94}, for a thin sheet, changes
in $z$ can be ignored over the scattering volume.  First, note that
the flux $F(\lambda,\tilde{t})=L(\lambda,\tilde{t})/4\pi D^2$, then
\begin{equation} \label{sb4}
 F_{\rm SC}(\lambda,t)=\int Q_{\rm SC}(\lambda,a) \sigma_g
 F(\lambda,\tilde{t}) \Phi(\mu,\lambda,a)\frac{n_d(\mbox{\boldmath
 $r$},a)}{4\pi r^2}da \, d^3r. 
\end{equation}
Solving $\mu$ for $\tilde{t}$ and substituting $r=\sqrt{\rho^2 + z^2}$,
$$
 \tilde{t}=t-\frac{\sqrt{\rho^2+z^2}}{c}+\frac{z}{c}
$$
and for a given time $t$
\begin{equation} \label{sb5}
 d\tilde{t} = -\frac{\rho d\rho}{cr}.
\end{equation}
Consider the spatial part of equation (\ref{sb4}) by transforming the
integral to cylindrical polar coordinates
$$
 I=\int n_d(\mbox{\boldmath $r$},a) 
 \frac{F(\lambda,\tilde{t})}{4 \pi r^2}\rho d\rho \, dz \, d\phi.
$$
Substituting equation (\ref{sb5}) and letting $dz=\Delta
z$ and 
$F(\lambda)=\int F(\lambda,\tilde{t}) d\tilde{t}$,
$$
 I=\frac{c\Delta z}{4\pi r} F(\lambda) \int  n_d(\mbox{\boldmath $r$},a) d\phi.
$$

\citet[hereafter MRN]{MRN77} found exctinction of starlight through
many lines of 
sight to be well reproduced if the distribution of dust grains in the
ISM (for a given range of $a$) is given by 
\begin{equation} \label{sb8}
 dn_{gr} = C n_{\rm H} a^{-3.5} da
\end{equation}
where $n_{gr}(a)$ is the number density of grains with radius $\le a$,
$C$ is a normalization constant, and $n_{\rm H}$ is the number density
of H nuclei.  Since $dn_{gr}/da$ and $n_d(\mbox{\boldmath $r$},a)$ are
both the number density of grains between $(a,a+da)$, equation
(\ref{sb8}) can be rewriten $n_d(\mbox{\boldmath $r$},a)=n_{\rm
H}(\mbox{\boldmath $r$}) f(a)$ where the exact form of $f(a)$ will be
determined in \S\ref{form-dust}.

Light echoes are observed as extended flux distributions, therefore we
measure the surface brightness $B_{\rm SC}$ of scattered light.  Noting
that $F_{\rm SC}=\int B_{\rm SC}\, \rho \Delta\rho\,d\phi$ and 
$F_{\rm SC}(\lambda,t) = \int F_{\rm SC}(\lambda,t,\phi) d\phi$,
equation (\ref{sb4}) now becomes
\begin{equation} \label{sb6}
 B_{\rm SC}(\lambda,t,\phi)=F(\lambda) n_{\rm H}(\mbox{\boldmath $r$})
 \frac{c\Delta z}{4\pi r \rho \Delta \rho}
 S(\lambda,\mu)
\end{equation}
where the integrated scattering function 
\begin{equation} \label{sb7}
 S(\lambda,\mu) =\int Q_{\rm SC}(\lambda,a)\sigma_g\Phi(\mu,\lambda,a)f(a)da.
\end{equation}
Since the observed flux $F({\lambda})$ and area element $\rho\Delta\rho$
diminish as $D^{-2}$, equation (\ref{sb6}) for surface brightness is
distance independent, as expected.  If the light echo is close to the
line of sight to $O$, the echo should suffer roughly the same
extinction as the observed flux $F(\lambda)$, mitigated of course by
the potentially higher dust density directly surrounding the echo
location.  

\subsection{Light Echo Width \label{form-width}}

\begin{figure}\centering
\epsfig{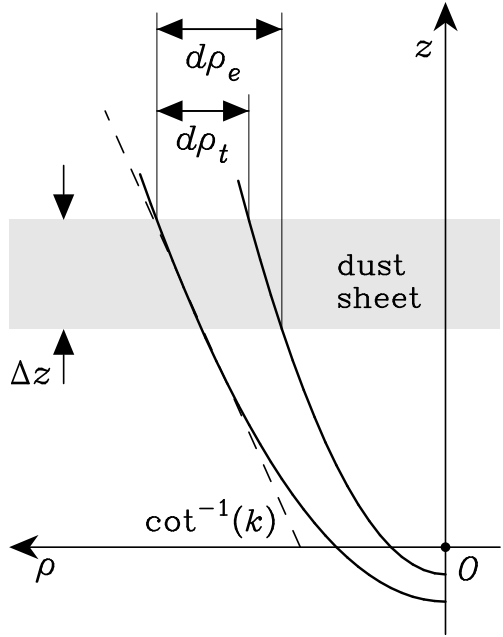}
\caption{Schematic showing the contribution of dust thickness and
  finite light-pulse duration on the observed width of a light
  echo. Adapted from \citet{Xu95}.
 \label{echowidth}}
\end{figure}

\begin{figure}\centering
\epsfig{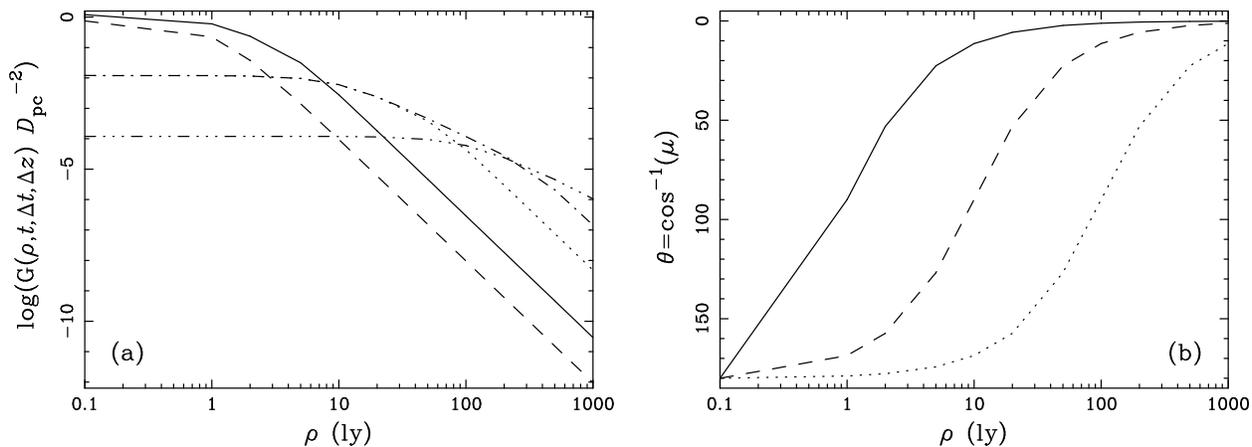}
\caption{Geometric factors in $B_{\rm SC}$ [eq.\ (\ref{sb6})].  (a)
The geometric function $G(\rho,t,\tau,\Delta z)$ [eq.\ (\ref{width4})]
for input $(t,\tau,\Delta z)$ as follows: solid line (1 y, 3 d, 0.1
ly); dashed (1 y, 0.25 y, 0.1 ly); dot-dashed (10 y, 3 d, 5 ly);
dotted (10 y, 0.25 y, 5 ly); dot-dot-dot-dashed (1000 y, 0.25 y, 25
ly).  (b) $\theta(\rho,t)$ for $t$=1 y (solid line), 10 y (dashed
line), 1000 y (dotted line).
\label{plg}}
\end{figure}

Since a light pulse has a finite duration $\tau$, the observed light
echo will have a width $d\rho_e$ resulting from the convolution of the
dust thickness $\Delta z$ and the pulse width 
\begin{equation}\label{width1}
 d\rho_t=\frac{cz+c^2t}{\rho}\tau.
\end{equation}
This is shown schematically in Figure \ref{echowidth}.  \citet{Xu95}
approximate the echo paraboloids in the neighborhood of $\Delta z$ by lines
of slope $k=d\rho/dz$, and find the convolution of two gaussians of FWHM
$\propto \Delta z$ and $dr_t$ yields
$$
 \frac{1}{2d\rho_e^2} = \frac{1}{2d\rho_t^2} - \frac{(k/2d\rho_t^2)^2}
  {(k^2/2d\rho_t^2) + (1/2\Delta z^2)}
$$
which simplifies to 
\begin{equation}\label{width2}
 \Delta z = \frac{\sqrt{d\rho_e^2-d\rho_t^2}}{k}.
\end{equation}
Using the light echo equation [eq.\ (\ref{sb1})] one finds
$$
 \frac{1}{k}=\frac{\rho}{ct}\,{\rm,\ and\ }\,
  d\rho_t=\left(\frac{\rho}{2t}+\frac{c^2t}{2\rho} \right)\tau
$$
with which equation (\ref{width2}), solved for $\Delta \rho=d\rho_e$,
becomes
\begin{equation}\label{width3}
 \Delta \rho = \sqrt{\left(\frac{\rho}{2t}+\frac{c^2t}{2\rho}\right)^2\tau^2
  +\left(\frac{ct}{\rho}\right)^2\Delta z^2 }.
\end{equation}
Since $z$ and $r=\sqrt{\rho^2+z^2}$ are both functions of
$(\rho,t)$, the geometric factors in equation (\ref{sb6}) 
\begin{equation}\label{width4}
 G(\rho,t,\tau,\Delta z) = \frac{c\Delta z}{4\pi r \rho \Delta \rho}
\end{equation}
and $\mu$ can be written in terms of observables $(\rho,t,\tau)$ with
$\Delta z$ the only unknown.  This allows an echo to be specified in
terms of the dust structure ($\Delta z$) rather than its observed
width ($\Delta\rho$).   The geometric factors $G$ and $\theta$
are shown in figure \ref{plg}.  In discussing the geometry it is often
easier to use units of years and light-years, since $c=1$. 

\subsection{Time-Integrated Flux \label{form-flux}}

To compute the surface brightess of an echo, one must know the
time-integrated monochromatic flux $F(\lambda)$.  If a homogenous
time-series of spectra exist for a given object, one 
simply integrates them over time, provided they have sufficient
spectral coverage for the wavelength range in question.  For this
paper, I wish to explore the observability of echoes around a variety
of objects, and over a large wavelength range (1000--11000\AA).  
Time-series of spectra are available for many (but not all) objects of
interest, and the wavelength coverage is highly inhomogenous, thus
such an approach for determining $F(\lambda)$ is not ideal.  Rather, I
propose a method by which a single spectrum can be used to approximate
the time-integrated flux.  This has the added bonus that it defines
the value of $\tau$ [eq.\ (\ref{width4})] in a consistent manner.  

\begin{figure}\centering
\epsfig{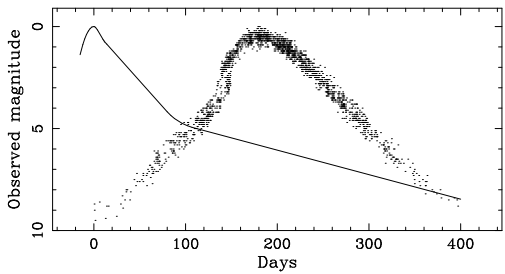}
\caption{Light curves for variable objects.  Solid curve -- mean light
  curve below maximum light in $B$ for Type II-L SNe.
  Points -- $V$ light curve (shifted by -5 mags) for the Mira $\chi$
  Cyg (from AAVSO International Database).  The maxima can be well
  represented by semi-logarithmic parabolae (\S\ref{const-exp}).
 \label{lc}}
\end{figure}

Although a light pulse has finite duration, its flux varies
substantially during $\tau$.  Furthermore, determination of the exact
value of $\tau$ is
ambiguous -- some commonly-used values are the time for flux to vary
by two or three magnitudes, or the ``effective width''
$\int{F(t)dt/F(t_{max})}$. 
Since fainter fluxes contribute less to light echo
signal, a compromise must be reached between inclusion of a reasonable
portion of the light curve (i.e.\ the wings), and exclusion of faint
fluxes whose contributions to an echo are negligeable compared to the
peak flux.  Sample light curves of composite Type II-P SNe
\citep{DB85} and the Mira star $\chi$ Cyg are shown in Figure
\ref{lc}.  These curves can be
approximated in the neighborhood of their 
maxima by parabolae in magnitude space.  A parabola with vertex
$(0,\log{f_0})$ and 
passing through $(t',\log{f_1})$ can be written
\begin{equation}\label{flux1}
 \log{f(t)}  =  \log{f_0} - \frac{t^2}{t'^2}L
\end{equation}
where $L=(\log{f_0}-\log{f_1})$.  The area $A$ under the curve between
$\pm t$ is found by integrating $f(t)$:
\begin{eqnarray}\label{flux2}
 A(t)=\frac{t'f_0\sqrt{\pi}}{\sqrt{L\ln{10}}}
 {\rm erf}(t\sqrt{L\ln{10}}/t').
\end{eqnarray}

The error function ${\rm erf}$ approaches unity if the operand is
$\gtrsim 1.4$, that is if $t>t_c=1.4t'/\sqrt{L \ln{10}}$. 
Let $t_n$ be the time required for the flux to
drop by $n$ magnitudes, or equivalently, for $\log[f(t)]$ to drop by
$0.4n$.  Applying this constraint to equation (\ref{flux1}), one finds
$t_n=t'\sqrt{0.4n/L}$.  Equation (\ref{flux2}) is roughly constant for
$t_n>t_c$ or $n\gtrsim 2.1$. Thus, a reasonable value of $\tau$ is
$2t_2$, i.e.\ 
the time required for the flux to rise then diminish by 2 magnitudes.  At
$\tau=2t_2$, equation (\ref{flux2}) reduces to
\begin{eqnarray}\label{flux3}
 A(t_2)=0.95\frac{f_0t'}{\sqrt{L}}\sqrt{\frac{\pi}{\ln{10}}}
  \approx 1.11\frac{f_0t'}{\sqrt{L}}.
\end{eqnarray}
By choosing to construct the parabola with $t'=t_2$,
$\sqrt{L}=\sqrt{0.8}$ and $A(t_2)=1.25f_0t_2$.  With this, an
approximation for the time-integrated monochromatic flux $F(\lambda)$ in
equation (\ref{sb6}) can be expressed by
$F(\lambda)=1.25F(\lambda,t_{max})t_2$, where $t_{max}$ is the time of
maximum light.   

How does this prescription for measuring $\tau$ compare with other
methods?  To address this, I have calculated a number of diagnostic
widths of the mean $B$ light curves for type Ia, II-P and II-L
supernovae from \citet{DB85}, which the authors provide in an
interpolated tabular form.  The results are given in Table
\ref{tbl-tau}.  The left columns give the values of $\tau$ for the
five methods noted below.  The right columns show how much of the
total flux of the light curve is included in this measurement of
$\tau$.  The methods are: (1) the effective width reported by
\citet{Spa94}; (2) the effective width I calculate integrated from
maximum light; (3) the effective width I calculate by integrating the
whole light curve; (4) the time for the light curve to actually rise
and fall by 2 mags from maximum light; (5) twice the time for the
light curve to fall 2 mags from maximum light.  There is a large
dispersion among the different methods, demonstrating the ambiguity of
this measurement.  The effective width measured from maximum light is
roughly half the value to rise/fall by 2 mags, showing that despite
the dispersion among methods, they are qualitatively consistent.
Aside from the type II-L curve, whose integrated flux saturates very
quickly, the effective-width measurements (methods 1--3) exclude
roughly one-quarter of the energy released in the outburst.

To some extent, it is up to the individual to select which
prescription for $\tau$ is preferred.  In the context of this paper,
the $t_2$ formulation presented above gives an estimate not only of
the width of the light pulse, which is needed in $G$ [eq.\
(\ref{width4})], but also of the time-integrated flux $F(\lambda)$ that
enters equation (\ref{sb6}).  For the purposes of exploring a general
parameter space, I will use equation (\ref{flux3}), and caution the reader
that integrated fluxes may be in error by 10--25\%.  In practice, the
spectral energy distribution of a variable object changes over $\tau$,
which this prescription for $F(\lambda)$ does not take into account.
Generating an empirical time-integrated spectrum is the obvious method
of choice, provided (as noted above) one has sufficient spectral and
temporal coverage.  In \S\ref{sec-93J}, I discuss the error associated
with this approximation applied to SN~1993J, and present an
easily-implemented method of correction.

\subsection{Dust Properties \label{form-dust}}

\begin{figure}\centering
\epsfig{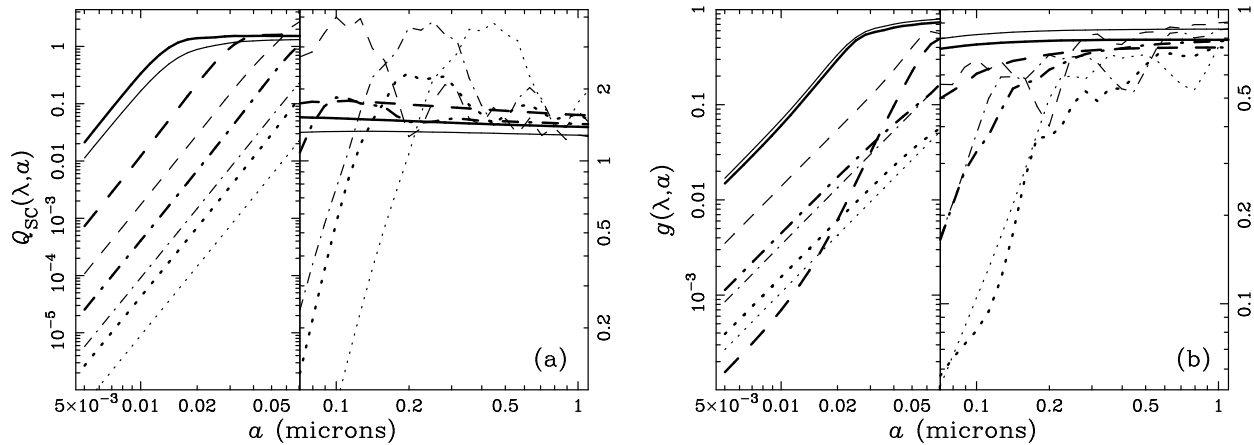}
\caption{(a) Scattering efficiency $Q_{\rm SC}(\lambda,a)$ and (b)
  forward scattering $g(\lambda,a)$ for carbonaceous (heavy lines) and
  silicate (thin lines) grains.  Solid lines show values for 1000\AA,
  dashed lines for 2500\AA, dot-dashed lines for 5000\AA, and dotted
  lines for 8900\AA.  For clarity, values for the smallest grain
  sizes are shown at different scales, as noted on the ordinates and
  abscissae.
\label{Qa}}
\end{figure}

In their classic analysis, {MRN} found that a mixture of carbonaceous
and silicate grains following a density distribution given by equation
(\ref{sb8}) adequately modeled observed starlight extinction.
\citet{DL84} and \citet{Lao93} calculated the dielectric functions for
graphite and ``astronomical silicate,'' from which they calculated the
dust properties ($Q_{\rm SC}$, $Q_{\rm ABS}$, and $g$) using the
dipole approximation (valid for $a/\lambda\lesssim 0.1$) and Mie
theory \citep{vdH57}.  $Q_{\rm ABS}$ is the absorption efficiency,
relevant when considering IR emission or optical extinction.
\citet[hereafter WD01]{WD01} have corrected an anomolous feature in
the silicate dielectric function, while \citet{Li01} have included the
effects of small PAH molecules in carbonaceous grains.  Values of
$Q_{\rm SC}(\lambda,a)$ and $g(\lambda,a)$ are
shown\footnote{Tabulated values are available at
http://www.astro.princeton.edu/$\sim$draine/dust/dust.diel.html} in
Figure \ref{Qa}, for both carbonaceous and silicate dust over a
variety of grain radii and wavelengths.

\begin{figure}\centering
\epsfig{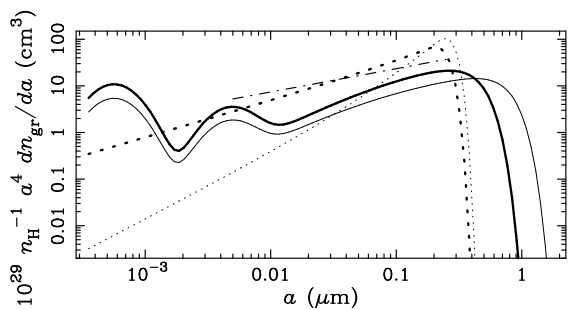}
\caption{Dust grain density distributions for carbonaceous (solid lines) and
  silicate (dotted lines) dust.  Thick lines are for the 
  $R_V=3.1,\ b_{\rm C}=6\times 10^{-5}$ model, and
  thin lines are for the $R_V=5.5,\ b_{\rm C}=3\times 10^{-5}$ model
  of {WD01}.  For clarity, and to allow the 
  reader to estimate the mass present in each size range, data are
  plotted in units of $a^4\,dn_{gr}/da$.  The dot-dashed line shows the
  corresponding density distribution from {MRN}.
\label{dnda}}
\end{figure}

{WD01} have constructed a new dust-grain density distribution
$f(a)$  fit to observed interstellar extinction.  I
refer the reader to their paper for the actual functional form, but
note that it contains a smooth cutoff for large and small grain sizes
and allows the slope $d\ln{n_{gr}}/d\ln{a}$ to vary, unlike the fixed
slope of equation (\ref{sb8}).  I have adopted their preferred Milky
Way dust model for \{case A, $R_V=3.1$, and $b_{\rm C}=6\times 10^{-5}$\},
shown in 
Figure \ref{dnda}, which works well at fitting both optical extinction
and infrared dust emission \citep{Li01} in diffuse clouds.  For
comparison, more reddened sightlines are well fit by the \{case A,
$R_V=5.5$, $b_{\rm C}=3\times 10^{-5}$\} density model, also shown in Figure 
\ref{dnda}.  The reddening ratio $R_V=A_V/E(B-V)$ \citep{Car89}, and
$b_{\rm C}$ is the 
total C abundance per H nucleus.  I caution that these particular
density distributions are not unique, and {WD01} discuss a
range of parameter space that fit the observed galactic optical
extinction (roughly) equally well.  Furthermore, extragalactic dust
densities need not follow that of our galaxy; the authors find a
different dust distribution for the LMC extinction, for example.
Uncertainty in the density distribution is to be expected, since e.g.\
(as noted by {WD01}) the dielectric functions are not
precise, dust grains have poorly-understood surface monolayers, and
grains are nonspheroidal.  As such, predictions based on the adopted
interstellar dust grain distribution should be taken with a proverbial
salt grain.

\subsection{Exposure Time Calculation \label{form-exp}}

Since the aim of this paper is to discuss light echo observability, I
briefly review the calculation of expsoure times for a generic
telescope, instrument, observing conditions and input spectrum.  The
signal-to-noise ($S/N$) ratio for an exposure time $t$ is given by the
CCD equation \citep{How00}
\begin{equation} \label{exp1}
S/N  = {C_{obj}t}/
 [( C_{obj} + C_{sky} + C_{bg})t +
     n_{DC}(t+t_{RO}) +
     n^2_{RD} + G^2\sigma^2_f]^{1/2}
\end{equation}
where $t_{RO}$ is the read-out time of the detector; $G$ is the CCD
gain [e$^{-}$ DN$^{-1}$]; 1 DN is one ``data number,'' or count (also
known as ADU); $n_{DC}$ is the dark current [e$^{-}$ s$^{-1}$
pix$^{-1}$]; $n_{RD}$ is the read noise [e$^{-}$]; $\sigma_f$ is the
digitization noise of the digital-to-analog converter [DN] and is
$\sim 0.3$ \citep{How00}; and $C$ indicates the count rate [e$^{-}$
s$^{-1}$ pix$^{-1}$], with $C_{obj}$ for the object observed,
$C_{sky}$ is that of the sky, and $C_{bg}$ is the background, e.g.\
nebulosity or galaxy surface brightness.  Dark current, read time,
gain and read noise are properties of the detector.  For a desired
$S/N$ it suffices to solve the quadratic in equation (\ref{exp1}) for
$t$.

The count rates are computed from the sensitivity of the system
\citep[as clearly explained in][]{STIS}, which gives the counts per
unit wavelength resulting from unit incident flux.  The sensitivity is
calculated as (the effective area of the telescope at $\lambda$)
divided by (the energy per photon at $\lambda$).  Let us define
$T_X(\lambda)$ to be the total throughput of the observational
configuration $X$ (e.g.\ telescope, filter, detector) at wavelength
$\lambda$, then sensitivity is ${T_X(\lambda) A \lambda}/{h
c}$ where $A$ is the area of the primary mirror, and $hc$ is Planck's
constant times the speed of light.  It follows that for an object with
surface brightness $B(\lambda)$, the count rate
\begin{equation} \label{exp2}
  C = \frac{A}{hc}m_X^2
  10^{-0.4A_X} \int B(\lambda) T_X(\lambda) \lambda d\lambda
\end{equation}
where $m_X$ is the platescale [arcsec pix$^{-1}$] and $A_X$ is the
extinction in filter $X$.  For a 2.4-m telescope (typical of a
moderate-sized ground-based observatory or {\em HST})
$A/hc=2.3\times10^{12}$ cm$^2$ erg$^{-1}$ \AA$^{-1}$.  Throughputs for
ACS, STIS and Johnson filters, including {\em HST} optics, are taken
from {\em synphot}.  Although a ground-based telescope will not have
the same throughput as {\em HST}, this approximation should be
sufficient.  For {\em HST}, detector characteristics and ``typical''
sky brightnesses ($m_V=22.7$ mag arcsec$^{-1}$, including zodiacal and
geocoronal light) are taken from the Instrument Handbooks
\citep{ACS,STIS}.  For a generic 2.4-m ground-based telescope, I adopt
sky brightness estimates for 3 days after new moon from \citet{Wal87},
and use characteristics typical of a SITE $2048\times 2048$-pixel
thinned and backside-illuminated CCD with $0.24\micron$ pixels, with
platescale computed at f/15.

For ground-based observations, I consider Johnson and Cousins filters
between $U$ and $I$.  STIS has three cameras of interest: the optical
long-pass F28x50LP filter and the two MAMA detectors (near and far
UV).  For ACS, I consider the optical Wide Field Camera (WFC), the
optical and near-UV High-Resolution Camera (HRC), and the UV
Solar-Blind Camera (SBC).  Characteristics for all filters and
detectors used in this work are listed in Tables \ref{tbl-filter} and
\ref{tbl-chip}, respectively.

\subsection{Extinction \label{form-ext}}

As noted in \S\ref{form-sb}, an echo and source should
suffer the same extinction, provided their angular separation is
small. Using the observed (i.e.\ not extinction corrected) source
spectrum for $F({\lambda})$, extinction need not enter equations
(\ref{sb6}) or (\ref{exp2}).  Since in this work I will often use template
spectra (\S\ref{model-spec}), extinction must be included in
surface-brightness calculations.  Furthermore, it is important to know
how much extinction is caused by the echoing dust,
in order to rule out unreasonable regions of the
parameter space (i.e.\ density or composition).  

The power per unit area of an electromagnetic wave traveling a
distance $L$ through dust of density $n_d$ will be attenuated by a
factor $\exp{(-n_dL C_{\rm EXT})}$ \citep{DL84}.  The factor 
$C_{\rm EXT}$ is
the extinction scattering cross section, defined as the sum of the
scattering and absorption cross sections $C_{\rm SC}$ and 
$C_{\rm ABS}$.  Since the cross section is related to efficiency by
$Q=C/\sigma_g$, the extinction (in magnitudes) can by calculated by
\begin{equation}
 A_\lambda=(2.5\log{e})N_{\rm H}\int{(Q_{\rm SC}+Q_{\rm ABS})\sigma_gf(a)da}
 \label{ext1}
\end{equation}
where $N_{\rm H}=n_{\rm H}L$ is the column density of hydrogen nuclei
along the optical path.  

I have integrated equation (\ref{ext1}) using the dust density,
scattering and absorption properties discussed in \S\ref{form-dust}.
Integration was performed using the extended Simpson's method
\citep{Press}, with values for $Q$ and $g$ interpolated
logarithmically in $\lambda$ and $a$.  Figure \ref{pleisf} shows the
resulting extinction for $R_V=3.1$ and 5.5 for Galactic dust, as well
as the individual contributions from silicate and carbonaceous
dust. For comparison, I also show the extinction curves from
\citet{Car89} in each panel.  As these are reported as
$A_\lambda/A_V$, I have normalized them to the computed total
extinction at 5495\AA\ (the value at which $A_\lambda/A_V=1$).  The
fit is very reasonable for both redenning parameters.  From these
curves, the value of $A_X$ in equation (\ref{exp2}) can be calculated,
given either an estimate for $A_V$ or $N_{\rm H}$.

\subsection{Non-Planar Dust Distributions \label{form-cse}}

Figure \ref{cse} shows schematic situations typical of a CSE around a
mass-losing star.  Mass-loss produces an $r^{-2}$ (gas) density
profile, shown here as a greyscale gradient.  The inner and outer
radii are chosen such that $R_{in}$ designates the radius at which
dust condenses in the wind, and $R_{out}$ is a matter boundary, such
as when the wind density reaches that of the ISM.  As discussed in
\S\ref{form-width}, one will observe an echo from all the material
located between the echo parabolae marking the beginning ($t_1$) and
end ($t_2$) of the light pulse or the star's high-flux state.  An echo
in a CSE differs from the simple case explored in the previous
subsections since density now varies as a function of scattering
position.  Some variable stars have periods larger than the
light-crossing time of the envelope, thus the envelope material swept
up between the echoes at $t_1$ and $t_2$ will be substantial, as shown
in panel (a).  In this ``thick echo'' regime, one may consider the
light echoing from a single column of width $\Delta\rho$ of this
envelope.  Although the overall geometry is not planar, one can apply
the planar approximation by integrating thin planar contributions along
this column.  If the period of variation is smaller than the envelope
size, one may consider the ``thin echo'' situation depicted in panel
(b), where a single sight line looks through many nested parabolae
(for clarity I have only drawn 12 echoes, although 39 fit into this
geometry).  The contributions from all such echoes must be included.


If the envelope has high optical depth, the single scattering
approximation is no longer valid.  For the purposes of this work,
single scattering will be sufficient, since $A_V<1$ in most dust
geometries (\S\ref{const-ext}).  A more thorough analysis of multiple
scatterings should be investigated for cases when $A_V\gtrsim0.3$,
however this first approximation will serve as a guide to whether such
echoes will be observable at all.

\section{THE LIGHT ECHO MODEL \label{sec-model}}

\subsection{Dust Scattering \label{model-dust}}

\begin{figure}\centering
\epsfig{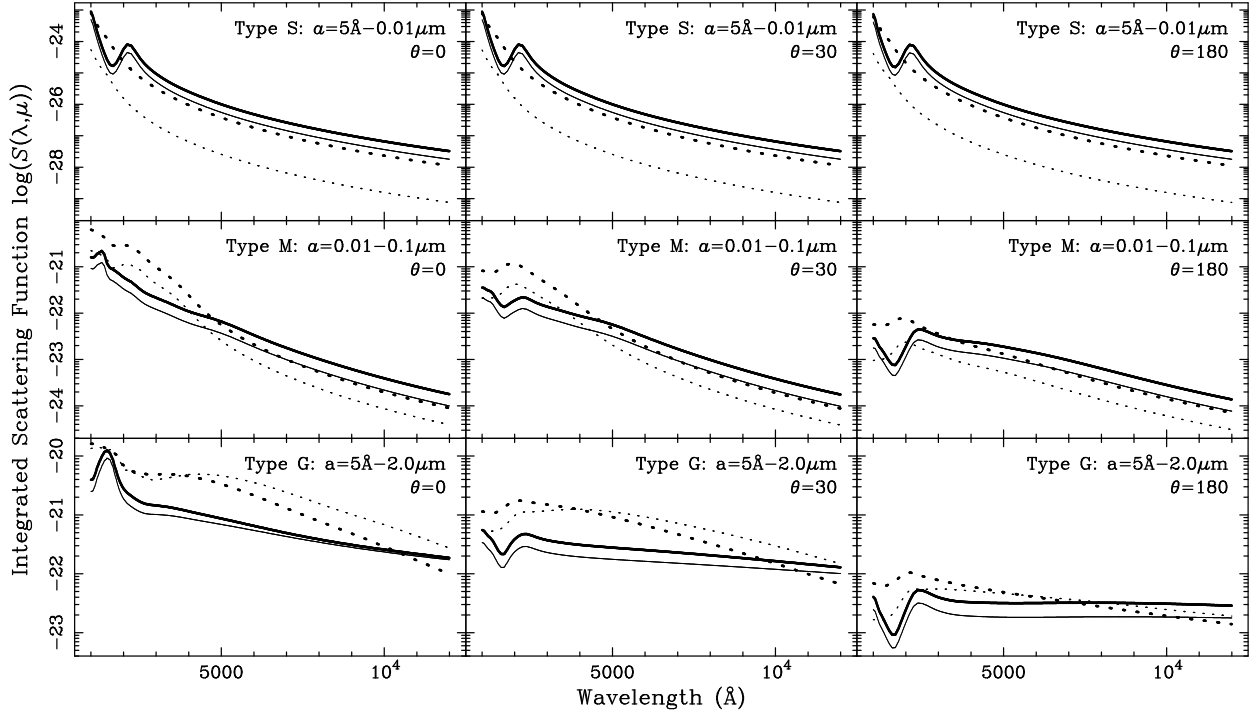}
\caption{Integrated scattering function $S(\lambda,\mu)$ [eq.\
(\ref{sb7})].  Line thickness designates the density distribution
model: thick lines are for $R_V=3.1$, and thin lines for $R_V=5.5$
(\S\ref{form-dust}).  Line style designates dust composition: solid
lines show carbonaceous dust, and dotted lines show silicates.  The
first row has been calculated for type S dust (\S\ref{model-dust}),
the second row for type M and the third row for type G.  The first
column is calculated for $\theta=0\degr$, the second for
$\theta=30\degr$, and the third for $\theta=180\degr$.
\label{plaisf}}
\end{figure}

Computation of the integrated scattering function $S(\lambda,\mu)$ is
a straightforward integral over grain sizes, using equations
(\ref{sb3}), (\ref{sb7}) and the dust-density models from {WD01}.
Selected results are plotted in Figure \ref{plaisf}, showing the
dependency of $S(\lambda,\mu)$ on density distribution model, grain
size range, composition, and scattering angle.  Columns show a given
scattering angle $\theta$ with varying grain-size range, and rows show
a fixed size range for varying $\theta$.  Changes along columns are
due strictly to the limits of integration for $S$, while changes along
rows are the result of variations in the scattering phase function
$\Phi$.  The three size-range distributions are denoted ``S'' (small,
$a=5\times 10^{-4} - 0.01$\micron), ``M'' (mid-sized, $a=0.01 -
0.1$\micron) and ``G'' (Galactic range, $a=5\times 10^{-4} -
1.0$\micron).  ``Galactic'' dust is the full density distribution
adopted by \citet{WD01}.

Scattering for grains with $a < 0.1\lambda / 2\pi$ occurs within the
Rayleigh limit, for which $S\propto\lambda^{-4}$.  This behavior is
clearly seen in the top row in Figure \ref{plaisf} and in the second
row, for $\lambda\gtrsim 5000$\AA.  For the smallest silicate grains,
$S\propto\lambda^{-4.3}$, while $S\propto\lambda^{-4.2}$ for
carbonaceous dust with $\lambda>3000$\AA.  This ``super-Rayleighan''
behavior reflects the dielectric function and resonance effects
(B. Draine 2002, private communication).  As the maximum grain-size
increases, scattering efficiency increasingly deviates from 
$\lambda^{-4}$ behavior, but UV scattering is generally more efficient
than optical.  

Of particular interest are the scattering peaks seen in the UV
(1000--2500\AA).  The presence of the strong 1500\AA\ carbonaceous
peak/dip in the bottom row (type G dust) demonstrates
the tendency of large dust to predominantly forward scatter near this
wavelength.  This conclusion is borne out by observing how the
scattering function changes with $\mu=\cos{\theta}$.  Regardless of
the adopted form for the scattering phase function, $\Phi$ should be
larger (smaller) for scattering toward (away from) the observer.  For
equation (\ref{sb3}), as $\mu\rightarrow \pm1$, $\Phi\rightarrow (1\pm
g)/(1\mp g)^2$, thus as the forward scattering efficiency $g$
approaches 100\%, $\Phi$ grows large for $\theta=0$ and
small for $\theta=180\degr$.  The degree of forward scatter $g\gtrsim
0.8$ for larger grains $(a\gtrsim 0.1\micron)$ at 1500\AA, but
$g\lesssim 0.04$ for grains smaller than $0.01\micron$.  Large grains
tend to forward-scatter UV light, while small grains scatter more
isotropically.  Thus we expect a UV bump near 1500\AA\ for $\theta\sim
0$ large-grain scattering, and a dip for $\theta \rightarrow
180\degr$, as observed in the bottom row of Figure \ref{plaisf}.

Maxima in $S$ near 1000\AA\ and 2175\AA\ in carbonaceous grains are
clear when the large-grain $\lambda=1500$\AA\ resonance is not present,
i.e.\ for the smallest grain sizes and for $\theta\rightarrow 180\degr$.
These two features demonstrate the high scattering efficiency
$Q_{\rm SC}$ of dust at short wavelengths.  The 
2175\AA\ bump is a known interstellar scattering and extinction band 
\citep{Ste65,SD65}, which {WD01} attribute to PAHs.  
The 1000\AA\ maximum is an artifact of the wavelength range plotted, and
$S$ continues to increase for shorter wavelengths (there is another
bump around 800\AA).  Large silicate grains also show a bump around
2000\AA, the origin of which is unclear.  

Figure \ref{plaisf} suggests that a great deal about dust can be
learned from carefully-planned observations of light echoes.  The
actual light echo flux will be a function of the total dust density,
grain size distribution, and the relative population of
carbonaceous-to-silicate grains, geometry and light-pulse
characteristics notwithstanding.  Since we generally know the latter
two, echo colors and magnitudes should constrain the dust properties.
Assume for now that the density distribution function is known, and
only the grain-size limits, density and composition are not.  The
observed echo spectrum will be a product of the dust scattering
$S(\lambda,\mu)$ and the pulse spectrum $F(\lambda)$
(\S\ref{model-spec}), from which object-specific echo models can be
created.

Echo geometry fixes $\Phi$, indicating on which column in Figure
\ref{plaisf} the total dust scattering efficiency must lie, while the
dust model can be chosen from $R_V$, as determined from stellar
spectrophotometry.  Optical colors alone may distinguish the largest
from the smallest dust grains (since smaller grains at longer
wavelengths follow the Rayleigh scattering relationship), however
since the dust number density (expressed here in terms of hydrogen
density) will likely be unknown, brightnesses themselves will not
conclusively distinguish between various particle-size scenarios.
Fortunately, blue to UV fluxes are a strong measure of dust size, and
UV-to-optical colors should lift the small-particle degeneracy.
Furthermore, aptly chosen wavebands can distinguish emission from the
various scattering maxima, which are sensitive to dust composition.
For example, the STIS MAMA far and near UV filters have average
wavelengths and FWHM of (1381\AA,324\AA) and (2310\AA,1237\AA),
respectively, with only mild overlap at 1500\AA.  The question remains
whether echoes at these wavelengths are easily observable.  Finally,
it is important to note that spectroscopy, even in low resolution but
covering a wide wavelength range, is likely the best observational
strategy for studying dust properties through light echoes.

\subsection{Variable Objects: Selection and Spectra \label{model-spec}}

\begin{figure}\centering
\epsfig{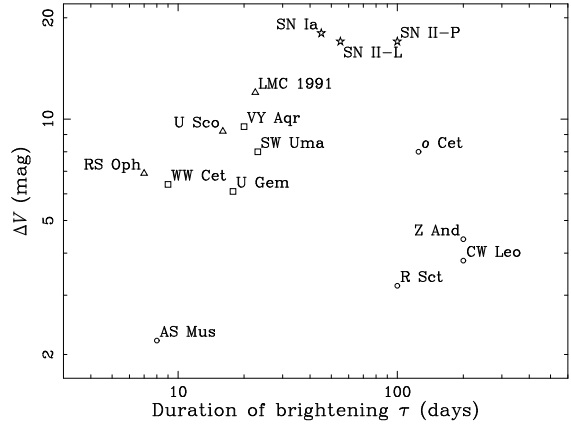}
\caption{Duration of brightening $\tau$ versus change in brightness
$\Delta V$ for the objects listed in Table \ref{tbl-var}  Supernovae
are marked with
stars, classical and recurrent novae with triangles, cataclysmic
variables (dwarf and TOAD novae) with squares, and mass-losing giant
stars with circles.  
\label{variables}}
\end{figure}

To predict the brightness of echoes from different sources, it is
first necessary to select candidate objects to study.  The primary
selection criterion is that the object vary in brightness by a
significant amount.  Without variability, any scattering material will
appear as a reflection nebula rather than as echoes, and no 3-D
information can be gleaned.  A second criterion is that the
duration of variability is short compared to the light-travel time
across at least one dimension of the scattering medium.   A burst of
duration 100 days can not give any spatial information about a dust
shell of radius 1 light day, but can still map the dust in a thin but
extended dust sheet since the sheet's dimensions are wider on the
plane of the sky than the pulse width.  
I list in Table \ref{tbl-var} classes of variable objects that
change in brightness by $\Delta V \gtrsim 2$ mag.  For easy
comparison, the data are also shown in graphical form in Figure
\ref{variables}.   Four broad categories are present:
supernovae, cataclysmic variables (CVs), pulsating giants and eruptive
stars. 

\begin{figure}\centering
\epsfig{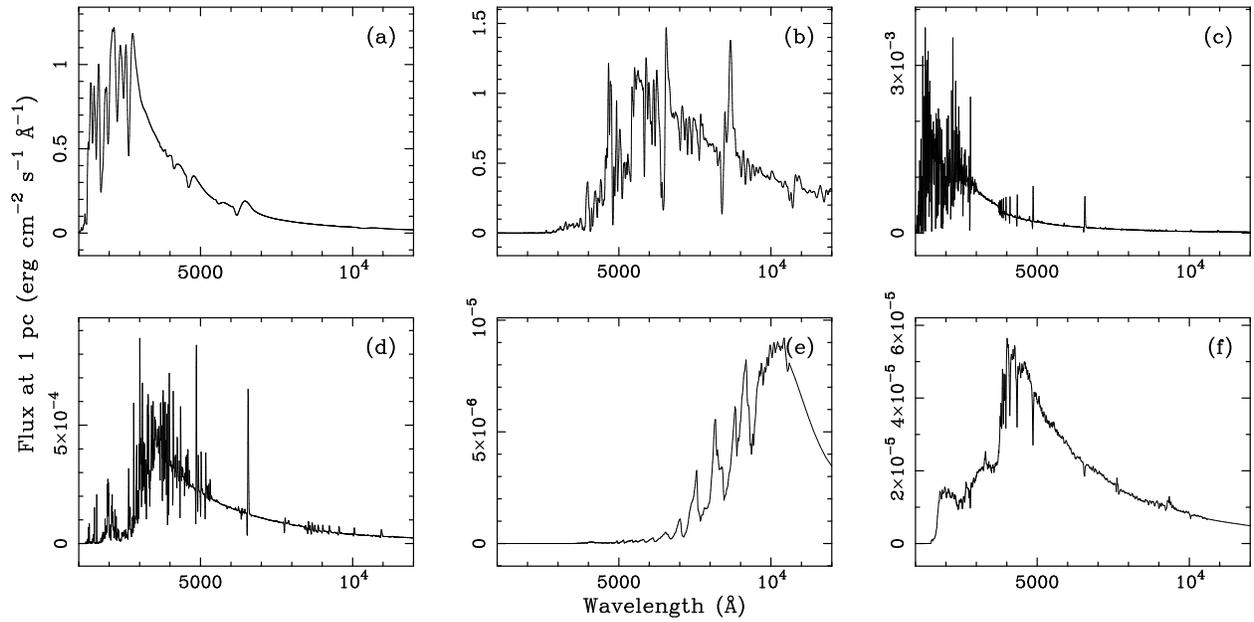}
\caption{Spectra used in this work, with fluxes normalized to a
distance of 1 pc.  {\em Phoenix} synthetic spectra for (a) SN~1987A at
day 1, and (b) at day 58, (c) a classical nova during the fireball
stage, and (d) a classical nova during the constant luminosity phase;
(e) the M7III star SW Vir (typical of
Miras) from the BPGS catalog
within {\em synphot}; (f) the F0IV star $\xi$ Ser 
from the BPGS catalog, normalized to the F6Ib spectrum of HD 8992
(typical of a bright Cepheid).
\label{plspec}}
\end{figure}

To model echo brightness, and to
discuss the relative brightness between different wavebands, it is necessary
to provide a realistic representation of the object's emission
spectrum in the factor $F(\lambda)$ in equation (\ref{sb6}). Selected
representational spectra for the first three 
categories are plotted in Figure \ref{plspec}, and
discussed below.  Each spectrum has been normalized to a distance of 1
pc.  This slightly-unconventional normalization has been chosen to
balance the conversion from physical area $(\rho\Delta\rho)^{-1}$ in $G$
(\S\ref{form-width}) to angular area.  Since these will represent the
flux at maximum brightness, I will refer to these as $F(\lambda,t_{max})$.

{\em Supernovae: }
Immediately following core collapse, type II supernova shock breakout is
brightest at shortest wavelengths.  The short-lived but intense pulse
of UV emission will photoionize any surrounding circumstellar
material, and likely evaporate the smaller, nearest ($d\ll1$ pc) dust grains
\citep[see][]{EC89}, however larger and more distant grains should
survive.  As shown in \S\ref{model-dust}, dust scattering is most
efficient in the UV for most echo geometries, making dust echoes of
the UV pulse an attractive observational target.  The later-time
spectrum is dominated by optical emission from cooling lines and
radioactive decay, and provides the longer-duration optical pulse that
is reflected in optical light echoes.  To model echoes from SNe, I use
synthetic spectra of SN~1987A from day 1 after core collapse (when the
SN was UV bright; Figure \ref{plspec}a), and day 58 after core
collapse (nearing the maximum optical light; Figure \ref{plspec}b),
generated with the {\em Phoenix} stellar and planetary atmosphere code
(E.~Baron 2002, private communication).  Both spectra were normalized using
a LMC distance of 50 kpc.  
A {\em Phoenix} synthetic spectrum of a Type
Ia event 25 days after explosion (not shown) has stronger flux than
SN~1987A (a Type II SN) at day 58, and while the microscopic features are
different, the overall spectrum is similar.  I will therefore use the
Type II spectrum for echo calculations, with a caution to the reader
that type Ia events are 1--2 mag brighter.

{\em CVs: } Within this category are classical, recurrent and dwarf
novae, and symbiotic stars, together spanning up to 10 mag between the
brightest and faintest outbursts, with such mechanisms as
thermonuclear runaway on an accreting white dwarf to accretion disk
instabilities.  To model echoes from classical novae, I use synthetic
spectra of a nova outburst during the fireball (Figure \ref{plspec}c)
and constant luminosity (Figure \ref{plspec}d) phases, again generated
with the {\em Phoenix} code (P.~Hauschildt 2002, private
communication).  Both have been normalized to Nova LMC 1991
\citep{Sch01} assuming an LMC distance of 50 kpc.  Recurrent novae
(including U Sco and RS Oph) should be adequately modeled as classical
novae with slightly less flux.  Dwarf novae (DNe) come in a variety of
flavors: U Gem, Z Cam, SS Cyg and SU Uma DNe have amplitudes of 2--6
magnitudes during normal outbursts, while TOADs -- SU Uma and WZ Sge
categories -- can have outburst amplitudes of 6--10 mag \citep{How95}.
As DNe are abundant, quasi-periodic, and can be bright, they may
provide a generous target list of echo candidates.  

DNe occur from instabilities in binary-star accretion disks.  The
theoretical spectrum of an infinitely-large, steady-state accretion
disk is given by $F_\lambda \propto\lambda^{-7/3}$ \citep{LB69},
however since the disk has a discrete inner edge (the accreting white
dwarf) this law is more reasonably written $F_\lambda
\propto\lambda^{-2.1}$ (J.\ Patterson 2002, private communication).  A
normalized (to 1pc) spectrum is easily generated from this law using
SS Cyg: $F(\lambda) = 1.4 \lambda^{-2.1}$ where $\lambda$ is in \AA\
\citep{PH84,Har99}.

{\em Pulsating Giants and Supergiants: } Pulsating giant stars include
(but are not limited to) long-period variables (LPVs) such as post-ABG
thermal-pulsating carbon stars and Miras, G--K type supergiant RV
Tauri stars, and F--K type Cepheids.  Figure \ref{plspec}e shows the
spectrum of the M7III star SW Vir, available in the
Bruzual-Persson-Gunn-Stryker (BPGS) spectrophotometry atlas within
{\em synphot}.  Itself a semi-irregular variable, SW Vir has a
spectral type similar to the prototypical Mira $o$ Cet.  This spectrum
has been normalized using the star's Hipparcos parallax of 7 mas
\citep{Per97}.  As expected of a cool giant, the energy distribution
is predominantly red.

A Cepheid has a type FIb spectrum during maximum brightness.  Many
such stars exist in the Jacoby-Hunter-Christian (JHC)
spectrophotometric atlas within {\em synphot}, but only between
3500--7400\AA.  To generate a Cepheid-like spectrum with the full
wavelength coverage needed, I scaled the F0IV spectrum of $\xi$ Ser
(from the BPGS catalog) to the F6Ib spectrum of HD 8992 (from the JHC
catalog), having first normalized HD 8992 to a Hipparcos distance of
0.82 mas, and corrected $\xi$ Ser for an extinction of $A_V=0.015$
(listed in the file header).  Figure \ref{plspec}f shows the resulting
spectrum.

{\em Eruptive Stars: } Many stars in this class do not satisfy the
required criterea, such as R Coronae Borealis, which fades rather than
brightens; UV Ceti stars, which have pulsation durations of seconds to
minutes; or Wolf Rayet stars, which only vary by $\Delta V\sim 0.1$
mag.  However, the S Doradus class is a high-luminosity blue star, typically
surrounded by diffuse nebulosity or expanding envelopes.  $\eta$ Car
is a famous member of this class, but with a period of over 5 years.
I note this class for completeness, and although echoes might be
observable in the extended envelopes, I will not make predictions for
these stars. 

\subsection{Observability of Light Echoes -- Constraints \label{model-const}}

Prior to integrating equation (\ref{sb6}), we
must specify the echo geometries and dust densities for each candidate
object.  Once specified, this integral yields the source   
count rate to be inserted into the CCD equation [eq.\
(\ref{exp1})], along with the 
appropriate values from Tables \ref{tbl-filter} and \ref{tbl-chip} and
an estimate of the background $C_{bg}$.   An echo is deemed
``observable'' if it can be detected at a given signal-to-noise level in a
``reasonable'' exposure time.  In the following sections, I discuss
estimates for exposure time, dust density and geometry, extinction,
and sources of background flux.  

\subsubsection{Exposure Time \label{const-exp}}   

Since light echoes are transient objects, they are effectively
detected by PSF-matched difference imaging, in which 
one matches the PSFs between two images using a variety of techniques
prior to subtracting them.  
This offers a
minimally-invasive method for removing all sources of constant flux,
with only sources of variable brightness remaining.  The differencing
technique of \citet{TC96}
has been shown by \citet[hereafter SC02]{SC02} to be quite effective
at detecting 
light echoes in {\em HST} imaging, as well as in e.g.\ the monitoring
of hot spots in SNR~1987A 
\citep{LSB00,Sug02}, detection of variable stars in globular clusters
\citep{Ugl99}, and the M31 microlensing project \citep{Cro92,Ugl02}.  

Direct photometric techniques (such as aperture photometry or even
PSF-fitting and removal) can only remove resolved point sources, but
have difficulty identifying faint surface-brightness features within
the fluctuations of an unresolved background component.  In
differencing two images, these unresolved sources are removed, leaving
(in principle) only statistical noise.  Even though this noise increases by
$\sqrt{2}$, the detection threshhold (i.e.\ required $S/N$) for an
echo decreases notably in a difference image, since we are looking for
coherent signal over 
a large, and now statistical-noise dominated, area.  A well-resolved echo
need only be detected at $S/N=2-3$ (per pixel) to appear unambiguously
above the difference-image background.  I will adopt a per-pixel
$S/N=3$ as the required detection threshold.

The assessment of a ``reasonable'' exposure time is subjective, but
not without practical constraints.  A reasonable series of
{\em HST} observations would image in 3--5 filters within perhaps double the
number of orbits.   With 50--70 minutes per orbit, a 30--90 minute
exposure seems a reasonable maximal exposure time.  Ground-based
observations are much more flexible, and I will consider a six-hour
(total) exposure time in any given filter as a representational (but
by no means fixed) upper limit.  

\subsubsection{Gas and Dust Density \label{const-gas}}   

For all model predictions, I will assume dust follows the type G (or
Galactic) dust distribution (\S\ref{model-dust}).  To calculate the
dust density, it is first necessary to specify $n_{\rm H}$, the
density of hydrogen nuclei, which normalizes $dn_{gr}/da$ to give the
dust density within interstellar gas.  The use of this normalization
factor depends on the assumption of a constant gas-to-dust ratio in
the ISM, however this same ratio may be unreasonable for circumstellar
environments.  The dust distribution integrates to yield a gas-to-dust
ratio of 124, and total number densities (per material) of $n_{\rm
C}=6.0\times 10^{-7}n_{\rm H}$ and $n_{\rm Si}=3.5\times 10^{-8}n_{\rm
H}$ for carbonaceous and silicate dust (respectively) with $R_V=3.1$,
while the $R_V=5.5$ distribution gives $n_{\rm C}=3.0\times
10^{-7}n_{\rm H}$ and $n_{\rm Si}=4.5\times 10^{-10}n_{\rm H}$.
For a given dust composition $(n_d=an_{\rm C}+bn_{\rm Si})$, the above
densities can be used to rescale $n_H$ to the appropriate dust content
within any environment.

Comparison of these dust densities with previously-reported values for
the ISM will be
problematic, since the {WD01} density profile favors a large
number of extremely small grains.  Small particles are inefficient
scatterers at optical wavelengths: $Q_{\rm SC}=10^{-10}-10^{-8}$ for
particles with $a=3\times 10^{-4}-10^{-3}$\micron, while $Q_{\rm
SC}>10^{-5}$ for $a>5\times 10^{-3}$\micron\ (Fig.\ \ref{Qa}).  
These grains contribute little to optical scattering, and have been
ignored in previous density distributions.  Their effect on the total
density is easily understood by considering the MRN distribution in
equation (\ref{sb8}), which integrates to yield 
$n_{gr}\propto(a_{min}^{-2.5})$ for $a_{min}\ll a_{max}$.  This
behavior strongly biases the density by their presence, yielding
values much larger than generally quoted.   For a
meaningful comparison with previous work, I have integrated the WD01
distribution over the size limits given by MRN: 
$0.005\le a_{\rm C} \le 0.25$ and $0.01\le a_{\rm Si} \le 0.25$.  
With these limits, the total number densities for $R_V=3.1$ are
$n_{\rm C}=7.6\times 10^{-11}n_{\rm H}$ and
$n_{\rm Si}=2.3\times 10^{-11}n_{\rm H}$, while for $R_V=5.5$, 
$n_{\rm C}=4.1\times 10^{-11}n_{\rm H}$ and
$n_{\rm Si}=2.8\times 10^{-12}n_{\rm H}$.

{\em Interstellar Dust:}
Following the multi-stage ISM model of \citet{MO77}, the majority of
interstellar gas will be in either the hot ($n_{d}\gtrsim 10^{-3}$)
or warm ($n_d\sim 0.1-1$) states.  Were we to regularly observe echoes from
this ambient-density medium, one would expect to see a diffuse halo
around echoing objects, rather than discrete arcs and rings.  This
suggests that we detect most light echoes from only the denser
regions of the ISM, however molecular clouds ($n_{\rm H}\gtrsim 1000$,
$r\sim50$pc)
have sufficiently high extinctions ($A_V\gtrsim5$) that echoes should
not be observed.  Rather, echoes likely originate from HI clouds with
$n_{\rm H}\sim 10$, which have sheet-like features and extinctions
$A_V<1$ (R.\ Chevalier 2003, Private Communication). 

{\em Mass-losing giants:} Post-AGB stars have mass-loss rates of
$10^{-8}$ to $10^{-4}$ M$_\sun$ yr$^{-1}$, yielding circumstellar
density profiles $n_d=K r^{-2}$ (where $K$ is a constant of
proportionality) , and with dust condensing within the outflow as near
as 8$R_*$, where $R_*$ is the stellar radius $\sim 3\times10^{13}$ cm
\citep{Hab96}.  \citet{Jus94} find gas-to-dust ratios of $\sim200$ for
oxygen-rich red-giant winds, which implies that a factor of $0.6n_{\rm
H}$ can be used to predict the dust density.  \citet{Vas93} find for
Miras a mass-loss scaling relation $\log{\dot{M}}=-11.4+0.0123P({\rm
days})$, yielding $\dot{M}\sim 10^{-9}$ $M_\sun$~yr$^{-1}$ for a 200
day Mira.
The equation of continuity predicts that the constant
$K=3\times10^{43}\dot{M}/v_{\infty}$ where $\dot{M}$ is in
$M_\sun$~yr$^{-1}$ and $v_\infty$ is the gas terminal velocity in
km~s$^{-1}$, typically 5--25 km~s$^{-1}$ for AGB stars, and 30
km~s$^{-1}$ for cool giants \citep{LC99}.  As such,
$K\sim3\times10^{33}$ cm$^{-1}$ for 200-day Miras.  Cepheids have mass-loss
rates between $10^{-10}$ to $10^{-6}$ M$_\sun$ yr$^{-1}$
\citep{Dea88}.  While most cepheids shed mass at the slower rate, high
mass-loss examples are known, such as RS Pup, whose mass-loss is
estimated by \citet{Dea88} to be $\sim10^{-6}$.  At this rate, the
density law will have $K=10^{36}$ cm$^{-1}$, while at
$\dot{M}=10^{-9}$, $K=10^{33}$ cm$^{-1}$.

{\em CVs:}  Dust can condense in a stellar wind once the equilibrium
temperature of the gas is less than $\sim1500$K, which occurs
\citep{LC99} at a radius 
\begin{equation}
r_c\simeq \frac{R_*}{2}\left(\frac{T_*}{T_d} \right)^\frac{4+p}{2}
\label{eq-dust}\end{equation}
where $p\sim1$ for silicate and $p\sim2$ for carbonaceous dust.  For a
dust condensation temperature $T_d\sim1000$K, $p=1.5$, a white
dwarf/accretion disk effective temperature of $(4-10)\times 10^4$ K,
and an inner radius $R_*=10^9$ cm, this yields $r_c=(1-16)\times
10^{13}$ cm.  UV resonance-line profiles from DNe imply mass loss
rates of $\dot{M}=10^{-9}$ M$_\sun$ yr$^{-1}$ \citep{Vit93} with
terminal velocities in the range of 3000--5000 km s$^{-1}$
\citep{War95}.  For the smaller-velocity limit, the continuity
equation gives $K=10^{31}$ cm$^{-1}$.

\subsubsection{Dust geometry \label{const-geo}}   

The geometric parameters entering $G$ [eq.\ (\ref{width4})] are listed in
Table \ref{tbl-geo} and discussed below. 

\begin{figure}\centering
\epsfig{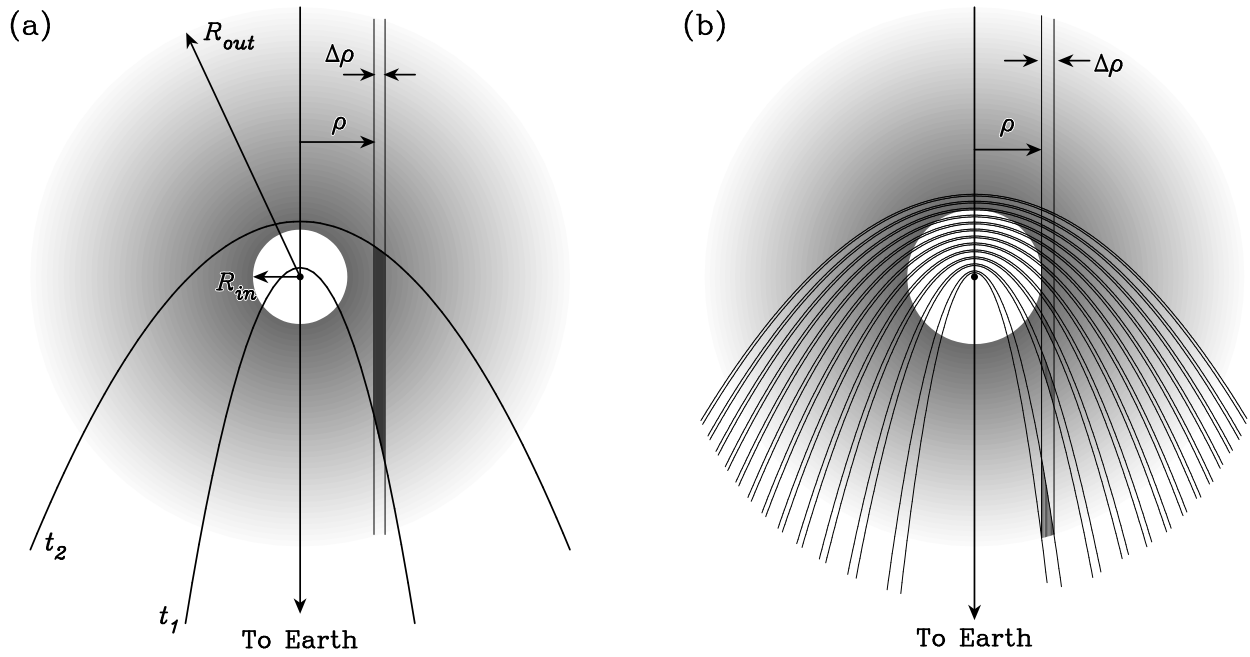}
\caption{Schematic diagrams of light echoes within CSEs.
(a) A thick-echo regime where only one echo passes through the CSE at
a time (typical of Miras).  (b) A thin-echo regime where many echoes
fill the CSE at once (typical of Cepheids).  See text
(\S\S\ref{form-cse}, \ref{const-ext} and \ref{const-bg}).
\label{cse}}
\end{figure}

{\em Mass-Losing Giant Stars:} Post-AGB and Mira stars (and
realistically, most stars late in their lifetimes, such as $\eta$ Car,
SN~1987A) have complex stellar outflows, often bipolar or asymmetric,
and are of critical interest as progenitors of an increasingly
complicated taxonomy of planetary nebulae \citep[as demonstrated
in][]{APN}. Although the scattering geometry is not planar in dusty
envelopes (as in Figure \ref{echogeom}), as discussed in
{\S\ref{form-cse}}, we can still apply the planar approximation.  As
shown in Figure \ref{cse}, the echo will occur between two parabolae
at $t_1$ and $t_2$, separated in time by $\tau$.  The choice of
geometric parameters representing typical echoes is somewhat
arbitrary.  Below, I discuss the geometries I will test.


The prototypical Mira $\chi$ Cyg has an inferred CO shell extending to
$\sim 3\times10^{16}$ cm \citep{JK98}, while the post-AGB carbon star
V Hya shows evidence of bipolar outflows at radii of $\sim 1.5\times
10^{16}$ cm \citep{Sug96}.  As the pulsation periods are large
compared to this size, I use the thick-echo approximation
(\S\ref{form-cse}).  As suggested by the aforementioned stars, one
radius to test is $3\times10^{16}$ cm, as well as
$10^{17}$ cm.  For both cases, I choose $t_1$ so an echo
passes through $r$ with $\theta=30\degr$.  I take the period to
be 200 days, $\tau$=50 days, and $\dot{M}=10^{-9}$ M$_\sun$ yr$^{-1}$.
This is shown schematically in Figure \ref{cse}a.

For Cepheids, I use an envelope geometry suggested by RS Pup.
Variations in light intensity in the surrouding nebula were
interpreted by \citet{Hav72} as arising from shells of reflecting
material spaced between inner and outer radii of roughly
$5\times10^{17}$ and $2.3\times10^{18}$ cm.  This inner radius is
roughly \onehalf ly, much longer than any Cepheid pulsation rate, thus
the thin-echo approximation applies.  I consider the innermost echo to
be located within a shell with radii given above, with $t_1$ such that
the inner echo passes through $R_{out}$ at $\rho=R_{in}$.  This is
shown to scale in Figure \ref{cse}b.  I test two cases: that for the
high mass-loss rate of RS Pup, and a more typical Cepheid mass-loss
rate of $10^{-9}$ M$_\sun$ yr$^{-1}$.

{\em Classical Novae:}
The first nova outburst will eject $10^{-4}-10^{-5}\ M_\sun$ of
material at hundreds to thousands of km s$^{-1}$ \citep{BE89},
expanding into the ISM in a manner analogous to that of a SN, the
physics of which is well studied \citep[e.g.][]{Che74}.  Indeed, this
is believed to be the case for Nova Per 1901 \citep{Sea89}.  Light
echoes from this event may highlight the CSE into which the ejecta
will travel, thereby allowing us to better understand (in the future)
how the ensuing nebula formed, as well as the mass outflow
characteristics of the star pre-nova (i.e.\ in its quiescent binary
period).  Recurrent novae may echo off their swept-up ejecta shells,
providing direct testing of predictions for these lower-energy
ejecta.  Furthermore, such echoes will constrain the recurrence time
of novae, a quantity which remains only speculative.  
Both types of bright novae may illuminate undisturbed ISM material as
well.  

Nova Per 1901 was a bright ($\Delta V\sim 13$) classical nova, and
notably it is the first object around which light echoes were
detected.  \citet{Sea89} have analyzed the archival imaging of GK Per
following its 1901 outburst, concluding that the echoes analyzed on
day 210 by \citet{Cou39} were from a plane of dust with $r=1.4$ pc at
nearest approach, and with thickness $\Delta z\sim0.15$ pc.  It is
unclear if this was true interstellar material, or a previously
swept-up shell of circumstellar material.  I will consider a sheet of
dust with these characteristics, as well as dust with (somewhat
arbitrary) thicknesses of $\Delta z=5$ ly at $z=$50, and $\Delta z=25$
ly at $z=250$ ly.

{\em Dwarf novae:} Being much less energetic in their outbursts than
classical novae, DNe may not evacuate their circumstellar material in
the same way, and hence may illuminate their quiescent stellar mass
loss.  GK Per currently has DN outbursts, the brighter of which might
echo off the swept up nova shell.  Of particular interest is the
geometry of outflows near the white dwarf.  While most accretion
systems drive narrowly-collimated jets (e.g.\ T Tauri stars, X-ray
binaries, quasars), non-magnetic CVs show evidence for mass loss with
large inferred opening angles \citep{Kni97}, suggesting these are
accretion systems in which jets are absent \citep{Kni98}.  I consider
a TOAD nova echoing from circumstellar ($r=0.025$ ly) and interstellar
($r=25$ ly) dust, as well as a dust sheet similar to that just
discussed from N Per 1901.  As the thickness of dust within the DN wind is
uncertain (due in part to the unknown orientation of the outflow axis)
I will arbitrarily take $\Delta z$ to be a light day.  

{\em Supernovae:}
Supernovae offer the means to illuminate the most distant material,
thereby probing the largest portions of the ISM. Unfortunately, the
observability of resolved echoes around SNe is limited by the low
frequency of nearby events.  
As in the previously-discussed objects, echoes from circumstellar and
interstellar dust offer
invaluable information on the progenitor's evolution and the
surrounding ISM.  In particular, echoes can probe multiple stages of
progenitor mass loss as imprinted in ``snow-plow'' contact-discontinuities
with the ISM \citep[e.g.,][for SN~1987A]{Cro91}.   I study six
possible dust geometries, two each from circumstellar, contact
discontinuity, and interstellar regions, using geometric parameters
similar to echoes observed around 
SN~1987A \citep{Cro95,Xu95}.

\subsubsection{Extinction \label{const-ext}}

\begin{figure}\centering
\epsfig{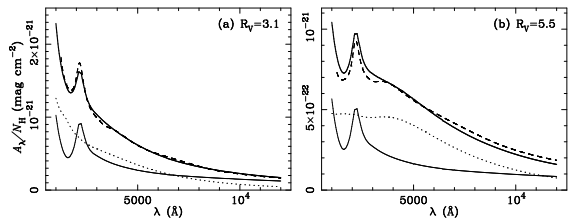}
\caption{Extinction curves as functions of $R_V$.  The lower curves
 show the extinction per unit hydrogen column density for carbonaceous
 (solid line) and silicate (dotted line) dust as integrated from
 equation (\ref{ext1}) (\S\ref{form-ext}).  The thick upper line is the
 total extinction, i.e.\ the sum of these two components.  The thin
 upper line is the extinction from \citet{Car89}, normalized so that
 $A_V=5.02\times10^{-22}$ for $R_V=3.1$, and $A_V=4.77\times10^{-22}$
 for $R_V=5.5$.
\label{pleisf}}
\end{figure}

To account for interstellar extinction from the light echo to Earth, I
use the extinction curve in Figure \ref{pleisf}, which relates
$A_{\lambda}$ in terms of $A_V$.  However, the choice of this
parameter depends on the particular line of sight and the distance of
the echo from Earth.  Rather than consider the myriad sightlines of
objects, I adopt $A_V=0.25$, with the caveat that predictions for
specific objects may require different extinction values.  

{\em Mass-Losing Giants:}  Consider the trajectory of
photons from the central star to the point of scattering and
subsequently out of the CSE.  In the single-scattering limit, light is
extincted along both paths.  An estimate of the extinction can calculated 
using the extinction curve in Figure \ref{pleisf}, from which 
$A_V=3\times10^{-22}N_{\rm H}$ for $R_V=3.1$ and a gas-to-dust ratio
of 200.
The column density is simply the integral of density along the photon
path $N_{\rm H}=\int n(r)dl$.  For a radial path between $R_{in}$ and
$r$, $N_{\rm H}=K(R_{in}^{-1}-r^{-1})$.  For a line-of-sight path
between $\theta$ and $\theta_{out}=\sin^{-1}{(\rho/R_{out})}$, and
with impact parameter $\rho$, $N_{\rm H}=K(\theta-\theta_{out})/\rho$.
$K$ is the 
constant of proportionality in the $r^{-2}$ density profile, discussed
in \S\ref{const-gas}, and is in the range $10^{33}-10^{37}$ cm$^{-1}$. 

The extinction in the radial trajectory from the star to the echo is
very sensitive to the radius at which dust condenses within the CSE,
since $N_{\rm H}\propto R_{in}^{-1}$ (provided $r\gg R_{in}$).
Using equation (\ref{eq-dust}), and assuming 
$p=1.5$ and that dust condenses around $T_d=1000$K,
$r_c=10R_*$ for a 3000K AGB star.  This is consistent with
\citet{Hab96}, who notes that for outflows from cool stars, dust
condenses at a radius of $8R_*$.  At this radius and with
$\dot{M}=10^{-9}$ M$_\sun$~yr$^{-1}$,  the maximal extinction is
$A_V=0.004$.
For a smaller and hotter Cepheid
such as RS Pup \citep[$R_*\sim185R_\sun$, $T_*\sim5600$K; ][]{Dea88},
dust will 
condense around $60R_*$. With
$\dot{M}=3\times10^{-6}$ M$_\sun$~yr$^{-1}$, 
$A_V\le0.4$, while for $\dot{M}=10^{-9}$ M$_\sun$~yr$^{-1}$,
$A_V\le4\times10^{-4}$.  By and large, the single-scattering
approximation is valid for CSE echoes since the extinction is $<1$.

\subsubsection{Background Flux \label{const-bg}}

{\em Mass-Losing Giants:} The entire column of gas containing the echo
(delimited with vertical lines) in Figure \ref{cse} will scatter
starlight into the line of sight, contributing the dominant source of
background light when the echo is imaged at time $t$.  Calculation of
this background requires an additional integral of equation
(\ref{sb6}) over the line-of-sight trajectory, also corrected for
internal extinction.  Consider panel (a) at a later time $t'$, say
when the variable star is passing through minimum and the echo has
left the shell.  The echo in the column shown will only be detectable
in a difference image if the total flux at $t$ (echo and background)
can be distinguished from the total flux at $t'$ (background
only).  
In the thin-echo approximation, there will never be a time when the
CSE does not have an echo in it.  Instead, one must ask whether the
entire column of echoes at $t$ is detectable against a background of 
echoes at a later time $t'$, such as half a period later.  To implement
this, I will use the column of flux at $t'$ as the background flux in
equation (\ref{exp1}).

{\em CVs and SNe:} I assume that any CSEs these stars might have are
optically thin compared to the mass-losing giants, and that extinction
effects can be ignored.  In this case, sources of background flux will
be the star's quescent light scattered off the surrounding dust, and
any unresolved background surface-brightness component.  The former is
estimated as the echo flux diminished by $\Delta V$ for the object (as
colors change during outburst, this is only an approximation).  For
the latter, I adopt a background component of $\mu_V=18.5$ mag
arcsec$^{-1}$, which is typical of extragalactic disks, in particular
the disk of M81 at the location of SN~1993J \citep{SC02}.  To scale
this to other wavelengths, I use the template spectrum of an Sb galaxy
from the Kinney-Calzetti catalog within {\em synphot}, normalized so
the flux in $V$ is unity.

\subsection{Observability of Light Echoes -- Predictions \label{model-pred}}

\begin{figure}\centering
\epsfig{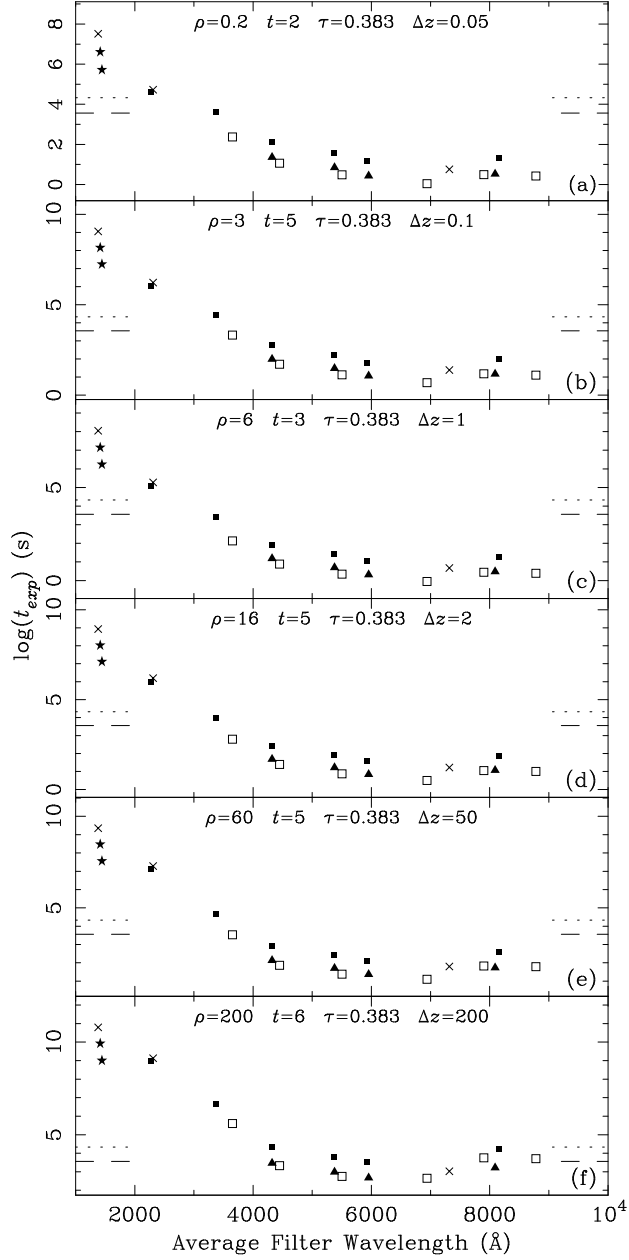}
\caption{Exposure times necessary to image an echo with $S/N=3$ (per
pixel) for a supernova.  Filters are plotted at their central
wavelengths (Table \ref{tbl-filter}), with STIS filters denoted by
crosses, Johnson/Cousins by open squares, and ACS filters by filled
symbols, with stars for the SBC, squares for the HRC, and triangles
for the WFC.  Geometric parameters are listed at the top of each panel
and correspond to those in Table \ref{tbl-geo}.  The dust is assumed
to follow a type-G (Galactic) size distribution.  Dashed and dotted lines
denote ``reasonable'' exposure-time upper limits for {\em HST} and
ground-based observations, respectively. UV filter
($\langle\lambda\rangle < 2600$\AA) exposure times are for the UV
burst spectrum, all others are for the optical peak spectrum.
\label{t87A}}
\end{figure}

\begin{figure}\centering
\epsfig{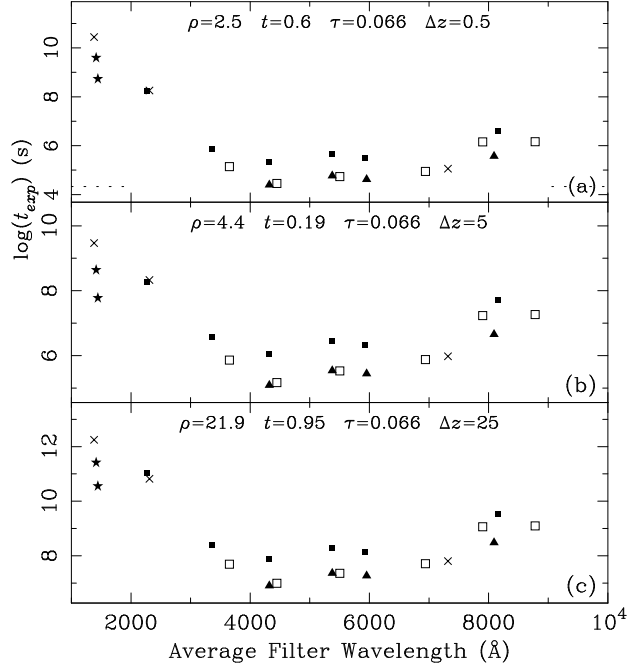}
\caption{Same as Figure \ref{t87A} but for a classical nova.  As in
Figure \ref{t87A},  UV filter
($\langle\lambda\rangle < 2600$\AA) exposure times are for the UV
burst spectrum, all others are for the optical peak spectrum.
\label{tGKPer}}
\end{figure}

\begin{figure}\centering
\epsfig{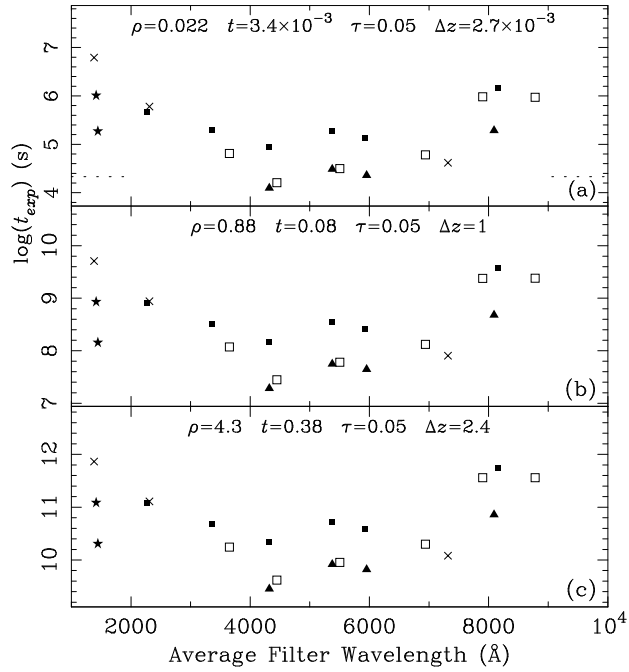}
\caption{Same as Figure \ref{t87A} but for a TOAD nova.
There is no distinction between UV and optical exposure times.
\label{tSSCyg}}
\end{figure}

\begin{figure}\centering
\epsfig{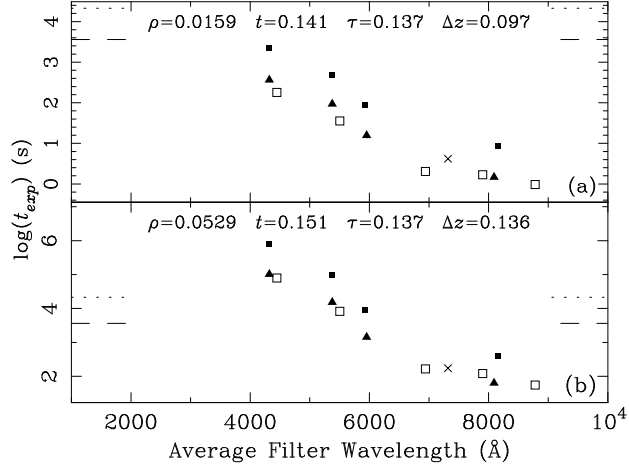}
\caption{Same as Figure \ref{t87A} but for a Mira, assuming the
thick-echo approximation (\S\ref{form-cse}).  UV exposure times
are all longer than the ordinate limits.
\label{tVHya}}
\end{figure}

\begin{figure}\centering
\epsfig{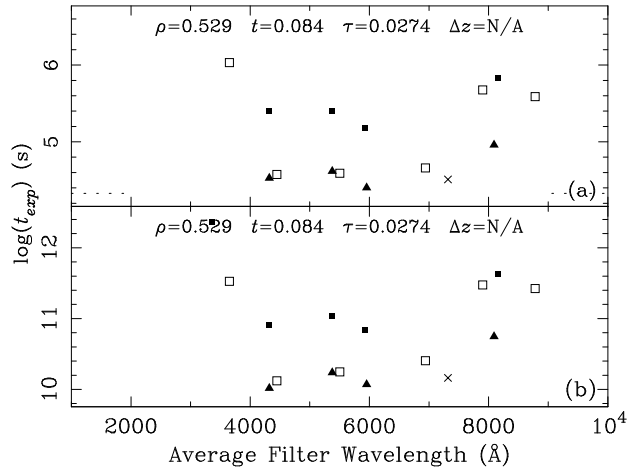}
\caption{Same as Figure \ref{t87A} but for a Cepheid, assuming the
thin-echo approximation (\S\ref{form-cse}).  UV exposure times are all
longer than the ordinate limits.  (a) is for a mass-loss rate of
$10^{-6}$~M$_\sun$~yr$^{-1}$, and (b) is for
$10^{-9}$~M$_\sun$~yr$^{-1}$.
\label{tRSPup}}
\end{figure}

Armed with scattering efficiencies $S$ (\S\ref{model-dust}), echo
constraints (\S\ref{model-const}), and light-pulse spectra
(\S\ref{model-spec}), calculating the expected count rate $C_{obj}$ of
light echoes is a simple integration of equation (\ref{exp2}) over
wavelength for a given set of filters and detectors.  The resulting
exposure times are plotted in Figures \ref{t87A}--\ref{tRSPup} for a
supernova, classical nova, TOAD nova, Mira, and Cepheid
(respectively).  From \S\ref{const-exp}, ``reasonable'' exposure times are
of order 1h for {\em HST} and 6h for ground-based imaging,
corresponding to $\log{t}$ of 3.5 and 4.3, respectively.  These have
been marked in each figure.

\subsubsection{Supernovae \label{pred-87A}}
Optical echoes from supernovae are known to be observable, and Figure
\ref{t87A} confirms this for a wide variety of echo geometries.
Interstellar UV echoes ought to be very difficult to observe, as most
lie above the exposure-time limits.  This is unfortunate since UV
echoes are the most effective at constraining the grain-size
distribution.  Since the input spectrum was for SN~1987A 1 day after
shock breakout, the UV pulse is likely stronger than the adopted
spectrum at short wavelengths.  This would decrease the exposure times
in the UV, perhaps making the near-UV echoes observable with long exposures.

\subsubsection{Novae \label{pred-GKPer}}
Light echoes from three novae have already been observed
(\S\ref{sec-intro}), however targeted searches have had difficulty
making positive detections \citep{vdB77,Sch88}.  We therefore expect
for these echoes to be fainter and require longer exposure times.
Figure \ref{tGKPer}a shows that circumstellar echoes may be observable
across most optical wavebands, but even the closer of the two
interstellar echoes [panels (b-c)] requires long integration times.
If interstellar echoes are to be observed, one is restricted to high
dust densities, energetic outbursts, and small depths ($z\lesssim 100$ ly).

\subsubsection{Dwarf Novae \label{pred-SSCyg}}
Exposure times for TOAD novae are shown in Figure \ref{tSSCyg}.
Circumstellar echoes from accretion-driven winds [panel (a)] may be
observable in optical wavebands for the brightest TOAD novae,
depending on the density and thickness of the gas.  If such echoes are
detected, either in ground-based imaging or with {\em HST}, we can
place much better constraints on the outflow properties.  Panel (b) is
calculated for the snow-plow of the wind into the ISM, and unless the
gas densities are unusually high, such echoes, as well as interstellar
echoes [panel (c)] will most likely not be detected.  Given that TOADs
are roughly 6 mag brighter than ordinary DNe, it is highly unlikely
that DNe will produce detectable echoes as well.

\subsubsection{Miras \label{pred-VHya}}
Figure \ref{tVHya} shows the exposure times for imaging an echo within
the CSE of a Mira.  These cool supergiants have extremely red spectra
and little UV flux, thus we expect their exposure times for the
shortest wavebands to be quite long.  As this is indeed the case,
filters shorter than Johnson $U$ are not plotted in this figure.
The results are encouraging in that a variety of echoes should be
observable.  
It is of interest however that while the template spectrum (Figure
\ref{plspec}f) has little flux blueward of 6000\AA, echoes from 
the $B$ and $V$ wavelengths may still be observable.  This is
fortunate since, as shown in \S\ref{model-dust}, larger 
wavelength coverage of an echo yields greater constraints on the
scattering dust.  As a matter of practical importance, these echoes
will be close (in angular separation) from the central source, which
itself is highly luminous and will thus create a confusion region.  If one
can obscure the bright central source (i.e.\ with a coronagraph),
light echoes can be observed within the CSE, provided the dust grain
sizes are sufficiently large.  

\subsubsection{Cepheids \label{pred-RSPup}}
Figure \ref{tRSPup} shows the exposure times to image variations in
nested, thin echoes within the circumstellar envelope of a bright
Cepheid.  As in the previous figure, filters shorter than Johnson $U$
are not plotted.  For two epochs separated by half a pulsation period,
such variations should just be detectable for the high mass-loss case
[panel (a)] with large platescale detectors.  They are effectively
undetectable for the lower mass-loss rates.  Modulations in the total
echo signal at a given sightline through the CSE will vary with the
same frequency as the pulsating star, which implies that a difference
of two images separated by half a period will appear as a series of
concentric bright and dark rings.  This is seen by \citet{Spa97} in
optical images of RS Pup, suggestive of the nested-echo
interpretation.  Of course, an additional challenge will be
disentangling the contribution of each nested echo.

\subsection{Caveats \label{pred-caveat}}

These results only delineate whether or not echo flux is
distinguishable from typical, smooth background noise
under limited conditions.  A
second criterium for observability is for the echo to be unambiguously
distinguishable from background sources.  Although in this work I
advocate the use of PSF-matched difference imaging, I must caution that this
technique is optimal only within inherent observational constraints,
which I briefly enumerate below.  (1) To correctly compute the
Fourier transform of a PSF, it must have sufficient spatial sampling,
which translates in practice to a FWHM$\gtrsim2.2$ pixels.  Most {\em
HST} and some good-seeing ground-based observations (with large
detector platescales) produce stellar profiles narrower than this
limit, necessitating image degradation and introducing subtraction
errors.  (2) Depending on the CCD used, a star that is bright
enough to fill a pixel beyond its linear regime can leave subtraction
residuals at the core, while saturated stars will create residuals
within 1--3 FWHM radii.  (3) Since point-sources have much higher flux
than the background, the increased Poisson noise will leave
subtraction residuals in the core of any bright (but linear) star.
(4) Bright point-sources in {\em HST} observations suffer from
large-radius diffraction patterns, particularly at short wavelengths.
In practice, these PSF-wings are never well subtracted by PSF-matching
(due to the much lower signal-to-noise in these regions), and
PSF-subtraction (i.e.\ with Tiny Tim model PSFs) is only moderately
effective at removing this signal at large radii.  (5) Many sources of
noise can take on stellar or extended profiles during data reduction.
Hot pixels and cosmic rays mimic undersampled stellar profiles
following geometric transformation.  (6) Charge-transfer efficiency
anomalies create extended tails around bright stars.  (7) Ghosts from
bright stars reflecting off optical surfaces produce a wide variety of
extended structure (e.g.\ WFPC2 ISR 95-06), including echo-like arcs.

Given these considerations, an echo should have a sufficiently large
angular separation from its source to not lie in the confusion region
of its PSF, i.e.\ within a few FWHM.  Stars on STIS images, for
example, have FWHM$\sim 0\farcs1$, while  ground-based
imaging can achieve $0\farcs5$ seeing under extremely-good
conditions. Let us consider radial separations 
larger than $0\farcs5$ to be safe from source confusion.  Most echoes
imaged to date have been rings or arcs of large angular extent
($\gtrsim 20\degr$ in position angle).  To reliably distinguish an
echo from a subtraction residual or anomalous background noise, it
should be sufficiently well-resolved to have an equally-large angular
size.  At $0\farcs5$ separation, an echo of arclength $30\degr$
will span roughly 6 pixels.  Echoes of larger
arclength should appear as coherant signal over the background and
through any other stellar PSFs.  The chance of observing of an echo is
maximized if we require the angular separation between the echo and
source to be greater than this lower limit of $0\farcs5$, which in
turn provides a maximum distance at which the echo source can be
located.  This parameter has been listed in the final column of Table
\ref{tbl-geo}, and the value can be easily scaled for any other
telescope and detector combinations.

The above arguments imply that the results presented in Figures
\ref{t87A}--\ref{tRSPup} only serve to suggest which echoes {\em could}
be detected.   The calculated exposure times do not account for the
various noise sources discussed, largely because they are much more
difficult to characterize in general terms.  Stellar subtraction
residuals are the dominant source of noise, and their presence is a
sensitive function of the wavelength observed as well as the density and
distribution of stars in the field.  The readership should fold
in these considerations with respect to their target fields.  Also, I
have attempted to generate generic dust geometries for producing
echoes, but the actual geometry and density can vary substantially
from target to target.  Finally, 
extremely short exposure times ($t_{exp}\sim1$ sec) suggest an echo should
be easily detected with an integration of typical duration (30--600s),
while very long exposure times imply an echo will only be observable
under very extenuating circumtsances.  

\section{Applications \label{sec-93J}}

The primary motivation for building the light echo model presented in
this paper was to quantify the observability of echoes around a
variety of variable objects.  Once echoes are detected however, this
same model is critical in interpreting the data.  Given an object's
object spectrum, either template or observed, one may construct a
series of expected spectral distributions against which the actual
echoes may be 
compared, to constrain the dust size distribution, composition and gas
density.  Similarly, if echoes are not detected around a given object,
this model provides limiting constraints on the above parameters.  

As an application of the techniques described in this work, I discuss
the light echoes recently discovered around SN~1993J
\citep[SC02]{Liu02}.  Both groups report the detection of an echo
south-west from the SN in WFPC2 imaging taken 8.2 years after core
collapse \citep[UT 19 April 1993;][]{Ben94}, at a radial distance from
the SN of roughly $1\farcs9$ and 120\degr\ in angular extent (denoted
SW770 in SC02).  From the echo equation [eq.\ (\ref{sb1})], this
corresponds to a depth $z=770$ ly from the SN, assuming a SN distance
of 3.63 Mpc \citep{Fre94}.  Using PSF-subtraction and PSF-matched
difference imaging, SC02 identify two additional echo candidates: one
located $1\farcs15$ north-east of the SN (denoted NE260), and a
marginal detection of SW770 in images taken only 1.8 years after core
collapse.  The direct and PSF-matched difference images in F555W from
SC02 are shown in Figures \ref{93J}a and \ref{93J}d, respectively.
The advantage of difference imaging is made most apparent by a
comparison of the noise reported by both groups in the F555W image.
\citet{Liu02} report a null detection at the position of NE260 with
measured noise of 9.2 DN pix$^{-1}$, while SC02 measure a flux of 4 DN
pix$^{-1}$ with a background noise of only 1.5 DN pix$^{-1}$.  The
reason the noise differs between the two groups is addressed in
\S\ref{const-exp}: difference-imaging removes surface brightness
fluctuations (from non-variable, unresolved stars, nebulosity, etc.)
that contribute to the ``background'' of direct photometric
techniques.

Having constructed a dust model from the the {MRN} grain-size
distribution and the scattering properties of \citet{DL84},
\citet{Liu02} present four model fits to their measured surface
brightness of SW770 in F450W, F555W and F814W.  Each model
overpredicts the blue photometry at $\sim 2\sigma$, while only the
pure-graphite model is consistent in the red.  The poor fitting in the
blue is most 
likely a result of their overly-dim measurement of SW770 in F450W.  If one
shifts the surface brightness in this filter to within the errors of
the F555W data (SC02 find roughly equal F450W and F555W fluxes), these
models are reasonable fits.  Of note, they suggest these echoes are
explained by dust with density $n_d\sim7\times10^{-10}$ grains cm$^{-3}$
embedded within gas of density $n_{\rm H}\sim3$ cm$^{-3}$.   How does
the echo model from this work compare?  

\begin{figure}\centering
\epsfig{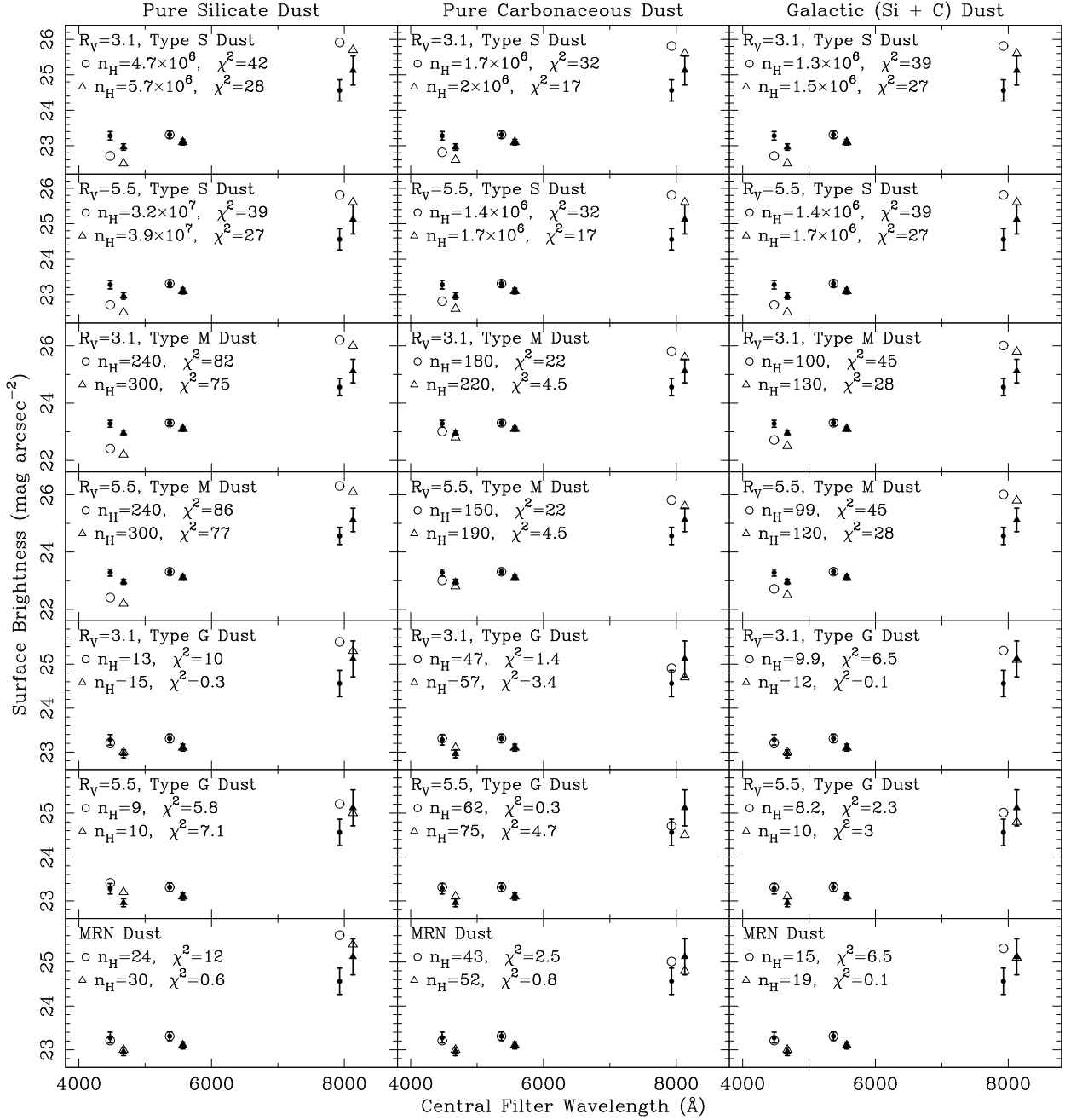}
\caption{Model predictions for the brightness of echoes around
SN~1993J for compositions as noted at the top of each column.  Solid
points are surface brigtnesses as measured by SC02 for the SW770 echo,
plotted about the central wavelengths of the corresponding WFPC2
filters (F450W, F555W, and F814W).  Open points are the model
predictions.   Within each pair, the left point (circles) 
is for the echo near PA 220, and the right point (triangles) is for PA
270.  Each model fit is scaled by $n_{\rm H}$ calculated to match the
F555W datum. 
These values of $n_{\rm H}$, as well as the $\chi^2$
residuals and the grain-size distribution, are noted in each panel.
\label{sb93J}}
\end{figure}

In Figure \ref{sb93J}, I present model surface brightnesses for a
variety of compositions and grain-size distributions discussed in this
paper.  In estimating the exposure times in \S\ref{model-pred}, I made
the assumption that the peak-flux spectrum of an object is a
reasonable representation of the spectrum over $t_2$, however for
detailed modeling of echoes, this may introduce systematic errors.
Integrated fluxes for SN~1993J exist only in ground-based filters
\citep{Ben94,Ric94}, however the echoes are measured in WFPC2 filters,
which differ enough from their Johnson and Cousins counterparts that
it may be unwise to directly intercompare them.  I generate the
time-integrated flux $F(\lambda)$ as follows.  

Predictions are based on an average of observed spectra from 17 and 24
days after core collapse \citep[][acquired from the SUSPECT supernova
spectrum archive]{Bar95}.  Although taken roughly 3 days before and
after maximum light, these particular spectra have wider wavelength
coverage than that taken at maximum light.  This spectrum ends around
9500\AA\ but the F814W WFPC2 filter extends to 11000\AA.  As an
approximate solution, I fill in this gap with the Type Ia
maximum-light template spectrum,  renormalized so the transition around
9500\AA\ is smooth.  I caution the reader that this will introduce
additional error into F814W predictions.  This composite spectrum is
integrated over the $B$, $V$, and $I_C$ response curves, and $t_2$ is
chosen to match the total ground-based fluence.  With $t_2=40$ days,
this spectrum reproduces the $V$ fluence to better than 1\%; note that
40 d is also the effective width ($\int{F(t)dt}/F(t_{max})$) of the SN
light curve reported in SC02.  The fluences in $B$ and $I_C$ are
correct using this value of $t_2$ if the $B$ flux is corrected by
0.94, and the $I_C$ by 0.77, when integrating over wavelength in
equation (\ref{exp2}).  These corrections are applied when computing
the predicted WFPC2 fluxes. 

Predicted surface brightnesses are compared to the observed values
reported by SC02 for two positions along SW770: the central region
around PA 220, and the extremal PA 270, at which the echo becomes
unusually blue.  In each panel, the gas density $n_{\rm H}$ has been
individually calculated to produce the observed F555W measurement for
each echo location.  I consider pure silicate dust, pure carbonaceous
dust, and a Galactic composition \citep[equal mixture of both as
given by][]{WD01}, for the type S, M and G grain-size distributions
with $R_V=3.1$ and 5.5.  Toward SN~1993J, \citet{Ric94} and
\citet{Pra95} find $E(B-V)=0.08-0.32$ mag, while values for $A_V$
range from 0.25--1.0 mag \citep{Ric94,Lew94}, yielding $R_V$ as high
as 7.3.  For completeness, I also present the model results for the
{MRN} size distribution.

As the model predictions in F814W are the least accurate, these data
should carry the least weight when considering the goodness of fit.  A
comparison of the fit to F450W and F555W only shows that all three MRN
models, and the Type G $R_V=3.1$ models for pure silicate and Galactic
dust, are the best fits.  It is of little surprise that the MRN fits
closely mimic the Type G $R_V=3.1$ fits, since both density models
were tuned to fit Galactic dust.  Including the F814W data as a 
constraint suggests the need for a highly carbonaceous dust
distribution.  

We can also judge the goodness of fit by the implied extinction of the
predicted dust distributions.  With $\tau=0.11$ y, $\rho=108$ ly, and
$\Delta\rho=11.5$ ly, equation (\ref{width3}) suggests the echoing
material has a line of sight depth of $\Delta z=47$ ly.  Using the
extinction data in Figure \ref{pleisf}, Type G silicate and
carbonaceous dust with $n_{\rm H}=10$ cm$^{-3}$ will produce
$A_V$=0.40 and 0.33 (respectively) for $R_V=3.1$, and 0.49 and 0.20
for $R_V=5.5$.  Thus, Galactic dust will produce respective
extinctions of $A_V$=0.73 and 0.69 for $R_V=3.1$ and 5.5.  Since
extinction scales linearly with hydrogen density, purely carbonaceous
type G dust would cause roughly 2 mags of visual extinction.  As also
noted by \citet{Liu02}, this is an unlikely fit as such optically
thick dust would obscure background star and galactic light as well.
Similarly, the type M and S dust models would require
unreasonably-high extinctions, and can be safely discarded.  With
$n_{\rm H}=10$ cm$^{-3}$, and including roughly 0.25 mag of Galactic
extinction, the Type G Galactic dust models predict $A_V\sim1$,
consistent with the extinction found near the SN by \citet{Ric94}.

It is of little surprise that the above results are inconclusive, since
there are many more degrees of freedom than constraints.  This
stresses the need for broad waveband coverage in the interpretation of
echo signal.  In particular, follow-up observations of SN~1993J should
integrate longer and include $R$ and $U$-band imaging.
Despite the shortcomings of this dataset, it illustrates a number of
important points. 

Foremost, the observed surface brightnesses are well reproduced by the
light echo model.  The dust in the disk of M81 (in the vicinity of
SN~1993J) appears to follow a grain-size distribution and have a
chemical composition similar to that
of Galactic dust.  Gas densities of order 10 cm$^{-3}$ are required to
produce the observed echoes, which for 
$R_V=3.1$ translates into dust densities of roughly $10^{-9}$
cm$^{-3}$, in agreement with that found by
\citet{Liu02}.  The color shift in PA between the two echo locations
discussed is easily explained by a small change in the density of the
echoing material.  This may be a simpler hypothesis than speculating
the presence of large compositional gradients of (mostly) pure
silicate to pure carbonaceous dust.  Furthermore, we may also propose
an explanation for the absence of the detected echo at other PAs.  The
noise reported by SC02 for their F555W difference image is 1 DN
pix$^{-1}$ for a scaled exposure time\footnote{Difference images are
empirically scaled to stellar flux of the reference image prior to
subtraction.  In this case, the input was a 2000 sec F555W WF4
integration taken 2001 June 4, and the reference was a 900 sec F555W
PC integration taken 1995 January 31.  A scaling of 953 sec includes the
effect of resampling the PC image to WF4 resolution.} of 953 s.  This
corresponds to 
a 1$\sigma$ detection limit of 25 mag arcsec$^{-2}$, or a hydrogen
density of about 2 cm$^{-3}$.  By no means conclusive, this is
suggestive that a simple density gradient in the cloud or sheet
containing the echoing dust can adequately explain the observations.
This substructure may be better understood through repeated
observations and with a more extensive wavelength coverage.

\section{Conclusions \label{sec-concl}}

In this paper, I have presented a model for calculating the brightness
of a scattered light echo.  I find that echoes from supernovae should
be the most easily observed, while cataclysmic variables may only
produce detectable echoes in the innermost circumstellar regions.
Post-AGB and Cepheid variables should also produce observable echo
signal, although the interpretation of such data will be technically
challenging.  

These results help explain why previous targeted searches have had
such difficulty in directly detecting light echoes.  For novae, it is
fairly likely that echoes were too faint to be observed.  For
supernovae, the case is not as clear.  In itself, a genuine null
result is interesting, since it places reasonably strong limits on the
dust density distribution along the line of sight.  More likely is
that many echoes were lost within the surface brightness fluctuations
of background unresolved starlight (for SNe in external galaxies) or
even moderately-crowded stellar fields, typical of observations close
to the Galactic plane.

\begin{figure}\centering
\epsfig{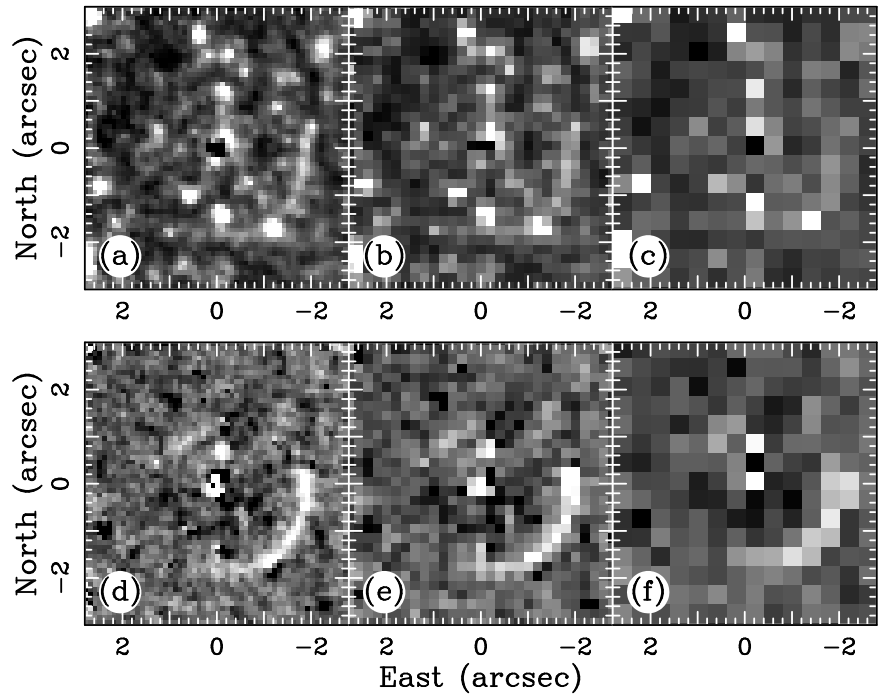}
\caption{F555W images of echoes around SN~1993J from {SC02}.  The SN
  (at the origin)   has been PSF subtracted but residuals remain.
  Panel (a) shows the direct image on WF4 taken in 2001, with resolution
  degraded by two in (b) and again in (c).  Panel (d) shows the
  PSF-matched difference between 2001 and 1995, and difference images
  between the input images degraded by two in (e) and again by two in (f).
\label{93J}}
\end{figure}

Consider the effect of simply increasing the platescale of the
detector in otherwise ideal observing conditions.  As a single pixel
images a larger portion of the sky, a light echo will fall on fewer
pixels, becomming increasingly confused with the background.  Although
the echo flux per pixel increases (aiding detection), so does the
background, and point sources become increasingly crowded.  The echo
SW770 around SN~1993J (\S\ref{sec-93J}, Fig.\ \ref{93J}a), happens to
be clearly visible above the background in direct F555W imaging
($m\simeq0\farcs1$ pix$^{-1}$), but is completely lost when the
resolution is degraded by a factor of four, as shown in Figure
\ref{93J}c.  A platescale of $0\farcs4$ pix$^{-1}$ is not uncommon by
ground-based standards, and for such an echo, direct photometric
techniques would likely have yielded a null detection.  Unlike SW770,
echo NE260 is only visible in the difference image, since in the
direct data, its flux is indistinguishable from background
fluctuations.  Even with the high resolution of {\em HST}, direct
imaging will miss echo signal.

Now consider the same exercise using PSF-matched difference imaging.
Since all sources of constant brightness will be removed, the echo
still stands out notably from the background when degraded images are
differenced (Fig.\ \ref{93J}e--f).  Still at issue is resolving the
echo from its source, which we see can not be done for NE260 at
$0\farcs4$ pix$^{-1}$.  A successful series of observations must
therefore have sufficient resolution to separate the echo from its
(variable) source, although this does not forcedly require the use of
{\em HST}.  In the case of SN~1987A, ground-based imaging was more
than sufficient to image interstellar echoes \citep{Xu95}.  For ideal
difference-image quality, effort should be taken to ensure a stable
and well-focused PSF, with individual exposures timed to avoid bright
star saturation, and total exposure time sufficient to minimize
background noise.

It is noteworthy that well-resolved echoes in ground-based observing
can be just as easily imaged as from space.  This is mainly due to the
platescale factor of $m_X^2$ in equation (\ref{exp2}).  A ground-based
integration will have $>9$ times the flux per pixel compared to the
ACS WFC camera, which more than compensates for the increased sky
background over vacuum observations.  Additionally, ground-based
observations generally have a wider field-of-view, thereby maximizing
the volume probed for light echoes.  However, increased platescale
corresponds to diminsished angular resolution, requiring the echo
source to be either closer to earth or more distant on the sky.  Also,
seeing greatly diminishes spatial resolution, thereby washing out any
substructure within the echo.  The observer must decide which
consideration is more important.

The question of {\em when} to observe an echo remains.  Unfortunately,
there is no concrete prescription.  The best image to difference
against is one in which there are no echoes, which implies either
observing the object prior to a burst, or median-combining a long
baseline of imaging to attempt to average out any echo signal.
Immediately following e.g.\ a nova outburst, the central source is too
bright for deep imaging, yet it is during this time that echoes sweep
through the innermost circumstellar material.  Coronagraphy is likely
the best options at these early times.  Long after an outburst, the
most easily observed echoes will come from forward-scattering
geometries, which for large $t$ imply large depth $z$ and angular
separation $\rho$.  However, since echo brightness decreases with
increasing dust distance and scattering angle, and suffers increasing
extinction with total optical path length, echoes will become fainter
as time progresses.  While only a small effect for recently discovered
SNe, this must be considered for Galactic SNe such as SN
1006.  Echoes from historic outbursts must be evaluated on a per-case
basis with the model presented here.  Outside these two extremes,
imaging should be taken in regular intervals of roughly twice the
outburst duration, to minimize the overlap of a single echo off the
same patch dust in both epochs.  With such considerations, light
echoes should be detected around a much broader list of objects than
currently exists.  To this end, an observational effort to put these
ideas into practice is currently underway for the four classes of
objects discussed in this work.

\acknowledgements

I offer my thanks to: Bruce Draine for making his dust scattering
calculations publically available, and for helpful discussions; Peter
Hauschildt and Ed Baron for generously providing synthetic Phoenix
spectra; Christian Knigge, Raghvendra Sahai, Roger Chevalier and Arlin
Crotts for useful discussions and feedback; and the referee, Bill
Sparks, for his critical reading of this manuscript, thoughtful and
insightful feedback, and for bringing echoes around RS Pup to my
attention.  This paper made use of publically-available data from the
AAVSO International Database, SUSPECT supernova spectrum archive,
General Catalog of Variable Stars, and the Catalog and Atlas of
Cataclysmic Variables.  {\em Synphot} is developed and maintained by
the Science Software Group at STScI.  This work was generously
supported by Arlin Crotts and STScI grants GO 8806, 8872, 9111, 9328
\& 9343, and NSF AST 02 06048.

\clearpage

\clearpage

\begin{deluxetable}{l c c c c c | c c c c}
\tablecaption{Mean Widths of Supernova Light Curves
\label{tbl-tau}}
\tablewidth{0pt}
\tablecolumns{10}
\tablehead{
\colhead{SN Type} & \multicolumn{5}{c}{$\tau$ (days)} & 
 \multicolumn{4}{c}{\% of total flux included} \\
\colhead{} & \colhead{1} & \colhead{2} & \colhead{3} & \colhead{4} &
\colhead{5} & \colhead{1,2} & \colhead{3} & \colhead{4} & \colhead{5}
}
\startdata
 Ia   & 24 & 22 & 31 & 38  & 46  & 65\% & 74\% & 80\% & 90\% \\
 II-P & 46 & 42 & 64 & 120 & 160 & 76\% & 65\% & 95\% & 95\% \\
 II-L & 26 & 28 & 39 & 60  & 72  & 94\% & 95\% & 95\% & 96\% \\
\enddata
\tablenotetext{1}{Effective width that
    \citet{Spa94} measured starting at maximum light.}
\tablenotetext{2}{Effective width that I measure starting at maximum light.} 
\tablenotetext{3}{Effective width that I measure using the entire
  light curve.} 
\tablenotetext{4}{Time for light curve to rise and fall by 2 mags.} 
\tablenotetext{5}{$2t_2$ as measured starting from maximum light.} 

\end{deluxetable}

\begin{deluxetable}{l c c c c c}
\tablecaption{Characteristics of Filters Used in This Work
\label{tbl-filter}}
\tablewidth{0pt}
\tablehead{
\colhead{Sysytem} &
\colhead{Filter} & \colhead{$\langle\lambda\rangle$\tablenotemark{a}} &
 \colhead{$\Delta\lambda$\tablenotemark{b}} &
\colhead{$T_{\rm peak}$\tablenotemark{c}} & 
\colhead{$C_{sky}$\tablenotemark{d}}
}
\startdata
STIS    &FUV MAMA & 1381.9 & 324.0 & 0.045 & $3.5\times 10^{-3}$\\
STIS	&NUV MAMA & 2309.5 & 1237. & 0.031 & $0.025$\\
STIS	&F28X50LP & 7318.7 & 2684.9 & 0.124 & $0.01$\\
ACS WFC & F435W & 4321.9 & 698.7 & 0.372 &  0.028  \\
ACS WFC & F555W & 5375.9 & 847.5 & 0.375 &  0.055  \\
ACS WFC & F606W & 5953.6 & 1575. & 0.468 &  0.136  \\
ACS WFC & F814W & 8092.2 & 1541. & 0.439 &  0.104  \\
ACS HRC & F220W & 2265.6 & 442.1 & 0.051 &  $10^{-4}$  \\
ACS HRC & F330W & 3366.2 & 410.7 & 0.105 &  $6\times 10^{-4}$  \\
ACS HRC & F435W & 4321.2 & 728.5 & 0.223 &  $4.3\times 10^{-3}$  \\
ACS HRC & F555W & 5369.6 & 841.4 & 0.239 &  $9.9\times 10^{-3}$  \\
ACS HRC & F606W & 5920.1 & 1557. & 0.276 &  0.024  \\
ACS HRC & F814W & 8152.3 & 1651. & 0.223 &  0.016  \\
ACS SBC & F115LP & 1415.0 & 354.5 & 0.0580 & 0.024   \\
ACS SBC & F125LP & 1445.3 & 334.1 & 0.0529 & 0.0027   \\
Johnson &$U$  	& 3673.0 & 478.0 & 0.050 & 0.17\\
Johnson &$B$  	& 4519.2 & 894.3 & 0.058 & 0.13\\
Johnson &$V$ 	& 5556.6 & 836.3 & 0.115 & 0.26\\
Johnson &$R$  	& 6784.6 & 1559. & 0.137 & 0.98\\
Cousins &$I_C$ 	& 7818.9 & 873.8 & 0.094 & 0.90\\
Johnson &$I$  	& 8285.7 & 1464. & 0.076 & 0.86\\
\enddata
\tablenotetext{a}{Average wavelength
$\langle\lambda\rangle=\int{T(\lambda)\lambda d\lambda}/
 \int{T(\lambda)d\lambda}$ in \AA.}
\tablenotetext{b}{FWHM of filter in \AA.}
\tablenotetext{c}{Peak throughput of filter and telescope.}
\tablenotetext{d}{Sky background in e$^{-}$ s$^{-1}$
pix$^{-1}$, for ``typical'' conditions defined in \S\ref{form-exp}.}
\end{deluxetable}

\begin{deluxetable}{l c c c c}
\tablecaption{Characteristics of Detectors Used in This Work
\label{tbl-chip}}
\tablewidth{0pt}
\tablehead{
\colhead{Instrument} & \colhead{Platescale} & \colhead{Read Noise} &
\colhead{Gain} & \colhead{Dark Current} \\
\colhead{} & \colhead{arcsec pix$^{-1}$} & \colhead{e$^{-}$} &
\colhead{e$^{-}$ DN$^{-1}$} & \colhead{e$^{-}$ s$^{-1}$ pix$^{-1}$}
}
\startdata
STIS FUV MAMA & 0.0246  & 0.   & 1.    & $3\times 10^{-5}$   \\
STIS NUV MAMA & 0.0246  & 0.   & 1.    & $2\times 10^{-3}$   \\
STIS CCD      & 0.0507  & 4.2  & 1.    & $2.5\times 10^{-3}$ \\
ACS WFC       & 0.05    & 5.0  & 1.0   & $2\times 10^{-3}$ \\
ACS HRC       & 0.027   & 4.7  & 2.2   & $2.5\times 10^{-3}$ \\
ACS SBC       & 0.032   & 0.0  & 1.0   & $1.2\times 10^{-5}$ \\
SITE f/15\tablenotemark{a} &
                0.15    & 5.4  & 3.3   & $4.\times 10^{-3}$  \\
\enddata
\tablenotetext{a}{Ground-based observations on a generic f/15 2.4m
  ground-based telescope.  See \S\ref{form-exp}} 
\end{deluxetable}

\begin{deluxetable}{l l c c c}
\tablecaption{Variable Objects with $\Delta V \gtrsim 2$ mag \label{tbl-var}}
\tablewidth{0pt}
\tablecolumns{5}
\tablehead{
 \colhead{Class} & \colhead{Example} & \colhead{$\Delta V$ (mag)} &
 \colhead{W (days)\tablenotemark{a}} & \colhead{Reference}
}
\startdata
\cutinhead{Supernovae}
SN Ia	  &  \nodata		& 18	&    40      & 1  \\     
SN II-P	  &  \nodata	   	& 17	&    140     & 1 \\	     
SN II-L	  &  \nodata	   	& 17	&    70      & 1 \\     
\cutinhead{CVs}
Cl Nova	  &  LMC 1991	   	& 12	&    20--25  & 2 \\     
U Sco 	  &  U Sco		& 9.2	&    16	     & 3\\     
RS Oph 	  &  RS Oph	   	& 6.9	&    6--8    & 3\\     
U Gem	  &  U Gem		& 6.1	&    17.8    & 3,4\\
Z Cam	  &  WW Cet	   	& 6.4	&    9	     & 3,4\\     
SS Cyg    &  SS Cyg		& 3.2   &    8       & 3,4\\
SU Uma	  &  SW Uma	   	& 8.	&    23.1    & 3,5\\
WZ Sge	  &  VY Aqr	   	& 9.5	&    20	     & 3,5\\     
Symbiotic &  Z And		& 4.4	&    200     & 6\\
\cutinhead{Pulsating Giants}
Mira/LPV  &  $\chi$ Cyg		& 8.	&    100     & 6\\     
RV Tauri  &  R Sct		& 3.2	&    100     & 6\\
Post-AGB  &  CW Leo	   	& 3.8	&    220     & 6\\     
Cepheid	  &  AS Mus	   	& 2.2	&    8	     & 6\\  
\cutinhead{Eruptive Stars}
S Doradus & $\eta$ Car & 9 & 2020 & 6 \\
\enddata
\tablenotetext{b}{Duration of outburst or $\tau$.}
\tablerefs{(1) composite light curves from \citet{DB85}; (2) \citet{Sch01};
(3) Catalog and Atlas of Cataclysmic Variables
(http://icarus.stsci.edu/$\sim$downes/cvcat/index.html);
(4) \citet{Ak02}; (5) \citet{How95}; 
(6) General Catalog of Variable
Stars (http://www.sai.msu.su/groups/cluster/gcvs/gcvs/) 
}
\end{deluxetable}

\begin{deluxetable}{c c c c c c c c c}
\tablecaption{Adopted Geometric Parameters
 \label{tbl-geo}}
\tablewidth{0pt}
\tablecolumns{9}
\tablehead{
 \colhead{Echo\tablenotemark{a}} &
 \colhead{$\rho$} & \colhead{$t$} &
 \colhead{$\tau$\tablenotemark{b}} & \colhead{$\Delta z$} &
 \colhead{$z$} & \colhead{$\theta$} & \colhead{$\Delta\rho$} &
 \colhead{$D_{max}$\tablenotemark{c}}\\
 \colhead{} &
 \colhead{(ly)} & \colhead{(y)} &
 \colhead{(y)} & \colhead{(ly)} & \colhead{(ly)} &
 \colhead{$(\degr)$} &   \colhead{(ly)} & \colhead{(kpc)}
}
\startdata
\cutinhead{Supernovae}
C         & 0.2 & 2.  & 0.003/0.383  & 0.05  & -0.99  & 168 & 2. & 25.\\
C         & 3.0 & 5.  & 0.003/0.383  & 0.10  & -1.6  & 118 & 0.47 & 380\\
B         & 6.0 & 3.  & 0.003/0.383  & 1.0  & 4.5  & 53. & 0.69 & 760\\
B         & 16. & 5.  & 0.003/0.383  & 2.0  & 23.1  & 35. & 0.92 & $2.0\times 10^3$\\
I         & 60. & 5.  & 0.003/0.383  &  50.  & 358  & 10. & 4.8 & $7.6\times 10^3$\\
I         & 200. & 6. & 0.003/0.383  &  200.  & 3330 & 3.42 & 8.8 & $25.\times 10^3$\\
\cutinhead{Novae}
C    & 2.5 & 0.6 & 0.003/0.066 & 0.5  & 4.9 & 27. & 0.19 & 320\\
I    & 4.4  & 0.19 & 0.003/0.066 & 5.0  & 50. & 5. & 0.80 & 560\\
I    & 21.9 & 0.95 & 0.003/0.066 & 25.0  & 250. & 5. & 1.3 & $2.7\times10^4$\\
\cutinhead{TOAD Novae}
C    & 0.022 & 0.0034 & 0.05 & 0.0027  & 0.069 & 30. & 0.17 & $0.3$\\
B    & 0.88  & 0.08   & 0.05 & 1.       & 5.     & 10. & 0.29 & 110\\
I    & 4.3   & 0.38   & 0.05 & 2.4      & 24.    & 10. & 0.30 & $10^3$\\
\cutinhead{Miras}
C   & 0.0159 & 0.141 & 0.137 & 0.097    & 0.027  & 30. & 0.002 & 2.0\\
C   & 0.0529 & 0.151 & 0.137 & 0.136    & 0.092  & 30. & 0.002 & 6.7\\
\cutinhead{Cepheids}
C   & 0.264 & 0.0977 & 0.0274 & N/A\tablenotemark({d}
                                        & N/A    & N/A & 0.05 & 33.\\
\enddata
\tablenotetext{a}{C: Circumstellar Echo; B: Contact Discontinuity; I:
 Interstellar Echo}
\tablenotetext{b}{If two values listed, the first is for the UV burst,
 the second for the optical maximum.  For Giant Stars, this column
 lists $\Delta\rho$ in ly.}
\tablenotetext{c}{Maximum distance of source assuming a mimimum
 echo/source separation of $0\farcs5$ using the {\em HST}.  See text
 (\S\ref{model-pred}).}
\tablenotetext{d}{For the thin-echo case, the sum of all echoes within
  the CSE is measured, thus $\Delta z$, $z$, and $\theta$ are not
  applicable.  }
\end{deluxetable}

%


\end{document}